\documentclass[preprint,aps,showpacs,floatfix,superscriptaddress,nofootinbib]{revtex4}
\usepackage[dvips]{graphicx}
\usepackage{epsf}
\usepackage{enumerate}
\usepackage{amsmath}
\usepackage{amssymb}
\usepackage{amsfonts}
\usepackage{latexsym}
\usepackage{revsymb}
\usepackage{bm}

\newcommand{\bra}[1]{\left\langle #1 \right|}
\newcommand{\ket}[1]{\left| #1 \right\rangle}

\renewcommand{\tilde}[1]{\overset{\lower1pt\hbox{$\scriptstyle{\sim}$}}{#1}}
\newcommand{\backtilde}[1]{\overset{\lower1pt\hbox{$\scriptstyle{\backsim}$}}{#1}}
\newcommand{\edlit}[1]{\overset{\lower1pt\hbox{$\scriptstyle{\backsim}$}}{#1}}

\newcommand {\bea} {\begin{eqnarray}}
\newcommand {\eea} {\end{eqnarray}}
\newcommand {\ba} {\begin{array}}
\newcommand {\ea} {\end{array}}

\def\veps{\varepsilon}

\def\l{\left}
\def\r{\right}
\begin{document}

\preprint{OCHA-PP-257}
\preprint{YITP-06-01}
\preprint{VPI--IPPAP--06--02}

\title{Matter Effect on Neutrino Oscillations from the violation of
Universality in Neutrino Neutral Current Interactions}
\author{Minako~Honda}\email{minako@hep.phys.ocha.ac.jp}
\affiliation{Physics Department, Ochanomizu Women's University, Tokyo 112-8610, Japan.}
\author{Naotoshi~Okamura}\email{okamura@yukawa.kyoto-u.ac.jp}
\affiliation{Yukawa Institute for Theoretical Physics, Kyoto University, Kyoto 606-8502, Japan}
\author{Tatsu~Takeuchi}\email{takeuchi@vt.edu}
\affiliation{Institute for Particle Physics and Astrophysics, Physics Department, Virginia Tech, Blacksburg VA 24061, USA}

\date{March 31, 2006}

\begin{abstract}
The violation of lepton-flavor-universality in the neutrino-$Z$ interactions can lead to 
extra matter effects on neutrino oscillations at high energies, beyond that due to the usual 
charged-current interaction of the electron-neutrino. 
We show that the dominant effect of the violation is a shift in the effective value of
$\theta_{23}$.  This is in contrast to the dominant effect of the charged-current 
interaction which shifts $\theta_{12}$ and $\theta_{13}$.
The shift in $\theta_{23}$ will be difficult to observe if the value of $\sin^2(2\theta_{23})$ 
is too close to one.  However, if the value of $\sin^2(2\theta_{23})$ is as small as $0.92$, then
a Fermilab$\rightarrow$Hyper-Kamiokande experiment can potentially place
a constraint on universality violation at the 1\% level after 5 years of data taking.
\end{abstract}


\maketitle
\section{Introduction}

When considering matter effects on neutrino oscillations, it is customary to
consider only the charged current interaction of 
the electron-neutrino mediated by $W$-exchange, 
and ignore the neutral current interactions of all three neutrino flavors 
mediated by $Z$-exchange.
This is because the universality of the neutral current interaction 
ensures that the phases acquired by the three neutrino flavors through
$Z$-exchange remain the same, and thereby do not lead to extra
mixing effects beyond that due to $W$-exchange.

However, in many models beyond the Standard Model (SM), the universality of
the $Z\nu_\ell\nu_\ell$ ($\ell=e,\mu,\tau$) couplings 
can be violated through radiative corrections, 
such as in Supersymmetric models with R-parity
violating interactions \cite{Lebedev:1999vc}, or through the mixing of the
light active neutrinos with heavy sterile ones \cite{Loinaz:2004qc}.
The existence of a $Z'$ which couples to the three lepton flavors 
differently can also mimic the violation of universality in $Z$-exchange \cite{Chang:2000xy}. 
Though the violation of $Z\nu_\ell\nu_\ell$ coupling universality in 
the particular models considered in 
Refs.~\cite{Lebedev:1999vc}, \cite{Loinaz:2004qc}, and \cite{Chang:2000xy}
are strongly constrained by 
the universality of the $W\ell\nu_\ell$ and $Z\ell\ell$ couplings,
they nevertheless
provide existence proofs that the universality of neutral current interactions 
cannot be taken for granted. 

The experimental bound on the violation of $Z\nu_\ell\nu_\ell$ coupling universality is
also very weak.  The sole constraint comes from CHARM and CHARM~II \cite{CHARM,PDB}:
\begin{eqnarray}
g^{\nu_e}   & = & 0.528 \pm 0.085 \;, \cr
g^{\nu_\mu} & = & 0.502 \pm 0.017 \;, \cr
g^{\nu_e}/g^{\nu_\mu} & = & 1.05^{+0.15}_{-0.18} \;=\; 0.87\sim 1.20 \;,
\label{CHARMconstraint}
\end{eqnarray}
where $g^{\nu_\ell}$ is the coupling of neutrino flavor $\nu_\ell$ to the $Z$,
normalized to 0.5 for the SM. 
These values were obtained from the measurements of the ratio
$R_\mu$ and the double ratio $R_e/R_\mu$, where
\begin{equation}
R_\ell \equiv
\dfrac{\sigma(\nu_\ell N \rightarrow \nu_\ell X)}
      {\sigma(\nu_\ell N \rightarrow \ell^{-} X)} \;.
\end{equation}
The constraint on $g^{\nu_\mu}$ was obtained from $R_\mu$, and the
constraint on the ratio $g^{\nu_e}/g^{\nu_\mu}$ was obtained from the
double ratio $R_e/R_\mu$ assuming charged current universality.
The constraint on $g^{\nu_e}$ was obtained from those on $g^{\nu_\mu}$ and 
$g^{\nu_e}/g^{\nu_\mu}$. 

As we can see from the above numbers, while $g^{\nu_\mu}$ is fairly well 
constrained to the SM value of 0.5, $g^{\nu_e}$ is ill constrained and can
deviate significantly from 0.5.
Of course, the sum of squares of the $Z\nu_\ell\nu_\ell$ couplings, namely
\begin{equation}
(g^{\nu_e})^2 + (g^{\nu_\mu})^2 + (g^{\nu_\tau})^2\;,
\label{InvisibleWidthConstraint}
\end{equation}
is well constrained to its SM value by the $Z$
invisible width measured by LEP and SLD \cite{LEPEWWG}, 
so any deviation in $g^{\nu_e}$ must
be accompanied by a corresponding deviation in $g^{\nu_\tau}$ to
maintain this agreement.  However, as long as $g^{\nu_e}$ and $g^{\nu_\tau}$
conspire to do so, large violations of universality are allowed.

In this paper, we investigate the effect of such violations of
neutral current universality on neutrino oscillations in matter.
If the violation is as large as that allowed by CHARM and CHARM II, then it could
lead to new effects that are measurable by long baseline neutrino oscillation
experiments.  If such effects are not seen, it could then improve upon 
the CHARM/CHARM II universality constraint.

This paper is organized as follows:  In section~II, we derive the effective potentials
due to the charged- and neutral-current interactions which
enter the effective Hamiltonians that govern neutrino and anti-neutrino propagation in matter.
In sections~III and IV, we approximately diagonalize the effective Hamiltonians for 
neutrino (III) and anti-neutrino (IV) propagation using the method of Ref.~\cite{HKOT},
and show how the effective mass-squared differences and effective mixing angles are
affected by the presence of neutral current universality violation.
In particular, we will show that the effective mass-squared differences are little affected, while
the shifts in the effective mixing angles are confined to just one angle; which angle this is depending 
on the mass hierarchy, and on whether the neutrino or anti-neutrino case is being considered.
In section~V, we discuss how these shifts in the effective mixing angles 
will manifest themselves in the neutrino and anti-neutrino oscillation probabilities, and 
point out that whether any effect can be seen or not depends crucially on the
value of $\sin^2(2\theta_{23})$ in vacuum.
In section~VI, we present the results of a numerical calculation of the effective mass-squared
differences, effective mixing angles, and oscillation probabilities, which validate the
approximations used in the previous sections.
In section~VII, we consider a hypothetical experiment in which the Fermilab NUMI beam 
\cite{NUMI,MINOS}
in its high-energy mode is aimed at a 1~Megaton class detector 9120~km away at Kamioka, Japan
(the planned Hyper-Kamiokande \cite{HyperK}) and discuss the potential constraint such an experiment
can place on neutral current universality violation.
Section~VIII concludes.

\section{The Effective Potentials due to $W$ and $Z$ exchange} 

Let us first derive the effective potentials for neutrino propagation in matter,
which account for the $W$- and $Z$-exchange interactions between the neutrino and
the matter fermions.
The effective potential due to $W$-exchange is well known \cite{MSW}, but we will re-derive it
in the following to provide a parallel to the $Z$-exchange case.

At momentum transfers much lower than the $W$ and $Z$ masses, the weak interaction
Hamiltonian of the neutrinos is given by
\begin{equation}
H_\mathrm{weak} = H_{CC} + H_{NC} \;,
\end{equation}
where $H_{CC}$ and $H_{NC}$ are the charged and neutral current contributions,
respectively:
\begin{eqnarray}
H_{CC}
& = &
\dfrac{G_F}{\sqrt{2}}
\left[ \bar{\nu}_\ell \gamma^\mu \left(1-\gamma_5\right) \ell
\right]
\left[ \bar{\ell}\gamma_\mu \left(1-\gamma_5\right) \nu_\ell
\right] \;,\cr
H_{NC}
& = & 
\rho\,\dfrac{G_F}{\sqrt{2}}
\left[ \bar{\nu}_\ell \gamma^\mu \left(1-\gamma_5\right) \nu_\ell
\right]
\left[ \bar{f} \gamma_\mu \left(g_V^{f}-g_A^{f}\gamma_5\right) f
\right] \;.
\label{eq:defweakH}
\end{eqnarray}
Here, $\ell$ is the lepton flavor ($\ell=e,\mu,\tau$);
$f$ denotes a generic fermion, and
$g_V^f$ and $g_A^f$ are its vector, and axial-vector couplings to the $Z$:
\begin{eqnarray}
 g_V^f & = & I_3^{f} - 2Q^{f} \sin^2\theta_W  \;,\cr
 g_A^f & = & I_3^{f} \;.
\end{eqnarray}
$I_3^{f}$ and $Q^f$ are, respectively, the isospin and electric charge of the fermion $f$.
If lepton-flavor-universality is violated in the neutral current interaction, 
the product $\rho G_F$ in the expression for $H_{NC}$ will depend on $\ell$.

After a Fierz transformation, $H_{CC}$ can be rewritten as
\begin{equation}
H_{CC} = 
\dfrac{G_F}{\sqrt{2}}
\left[ \bar{\ell}\gamma^\mu \left(1-\gamma_5\right) \ell
\right]
\left[ \bar{\nu}_\ell \gamma_\mu \left(1-\gamma_5\right) \nu_\ell
\right] \;.
\end{equation}
The forward scattering amplitude of a neutrino $\nu_\ell$ against a 
non-relativistic lepton $\ell$ via $W$-exchange is then
\begin{eqnarray}
\mathcal{M}_{CC} 
& = & 
\dfrac{G_F}{\sqrt{2}}
\bra{\ell}\bar{\ell}\gamma^0\left(1-\gamma_5\right)\ell\ket{\ell}
\bra{\nu_\ell}\bar{\nu}_\ell\gamma_0\left(1-\gamma_5\right)\nu_\ell\ket{\nu_\ell} \cr
& = &
\sqrt{2}\,G_F
\bra{\ell}\ell^\dagger\ell\ket{\ell}
\bra{\nu_\ell}\nu_\ell^\dagger\left(\frac{1-\gamma_5}{2}\right)\nu_\ell\ket{\nu_\ell} \cr
& = &
\sqrt{2}\,G_F N_\ell
\left( \phi_{\nu_\ell}^\dagger \phi_{\nu_\ell} \right) \;,
\end{eqnarray}
where $N_\ell \equiv \bra{\ell}\ell^\dagger\ell\ket{\ell}$ is the density of the charged lepton $\ell$, and 
$\phi_{\nu_\ell}$ is the two-component wave-function of the left-handed neutrino $\nu_\ell$. 
This shows that the effective potential that the neutrino experiences as it travels through
matter is
\begin{equation}
V_{CC} = \sqrt{2}\, G_F N_\ell\;.
\end{equation}
In ordinary matter, $N_\mu=N_\tau=0$. Therefore,
\begin{equation}
V_{CC} = 
\left\{
\begin{array}{ll}
\sqrt{2} G_F N_e & \qquad \mbox{(for $\nu_e$)}\,, \\
0                & \qquad \mbox{(for $\nu_\mu$, $\nu_\tau$)}\,.
\end{array}
\right.
\end{equation}

Similarly, the forward scattering amplitude due to $Z$-exchange between a neutrino and a
non-relativistic fermion $f$ is given by
\begin{eqnarray}
\mathcal{M}_{NC} 
& = & 
\dfrac{G_F}{\sqrt{2}}
\bra{f}\bar{f}\gamma^0\left(g_V^f-g_A^f\gamma_5\right)f\ket{f}
\bra{\nu_\ell}\bar{\nu}_\ell\gamma_0\left(1-\gamma_5\right)\nu_\ell\ket{\nu_\ell} \cr
& = &
\sqrt{2}\,G_F
\bra{f}g_V^f( f^\dagger f )\ket{f}
\bra{\nu_\ell}\nu_\ell^\dagger\left(\frac{1-\gamma_5}{2}\right)\nu_\ell\ket{\nu_\ell} \cr
& = &
\sqrt{2}\,G_F g_V^f N_f
\left( \phi_{\nu_\ell}^\dagger \phi_{\nu_\ell} \right) \;,
\end{eqnarray}
where we have set the $\rho$-parameter to one, and 
$N_f = \bra{f}f^\dagger f\ket{f}$ is the density of the fermion $f$.
The effective potential due to the neutral current interaction is then
\begin{equation}
V_{NC} 
= \sqrt{2}\, G_F g_V^f N_f
= \sqrt{2}\, G_F \left( I_3^f - 2Q^f \sin^2\theta_W \right) N_f \;.
\end{equation}
Since $N_e = N_p$ in electrically neutral matter, we find
\begin{eqnarray}
V_{NC} & = & 
\sqrt{2}\,G_F
\left[
 \left( -\dfrac{1}{2} +  2 s_W^2 \right) N_e
+\left(  \dfrac{1}{2} -  2 s_W^2 \right) N_p
+\left( -\dfrac{1}{2} \right) N_n
\right] \cr
& = &
-\dfrac{1}{2}\left(\sqrt{2}\, G_F N_n\right) \,.
\end{eqnarray}

Assuming $N \equiv N_e = N_p \approx N_n$,
which is valid for the lighter nuclei
which constitutes most of the Earth, we can relate $N\;(\mathrm{cm^{-3}})$ 
to the matter density $\rho\;(\mathrm{g/cm^3})$ via the Avogadro number $N_A$:
\begin{eqnarray}
2N = N_p + N_n = \rho N_A\;.
\end{eqnarray}
Then,
\begin{eqnarray}
\sqrt{2}\,G_F N
& = & \sqrt{2}\,\frac{G_F}{(\hbar c)^3} (\hbar c)^3 \frac{N_A}{2} \rho \cr
& = & \frac{1}{2}\,(7.6324\times 10^{-5}\,\mathrm{eV}^2)
      \times \left(\dfrac{ \rho }{ \mathrm{g/cm}^3 }\right)
      \times \left(\dfrac{ 1 }{ \mathrm{GeV} }\right)\,,
\end{eqnarray}
where we have used
$G_F/(\hbar c)^3 =1.16637(1)\times 10^{-5}\;\mathrm{GeV}^{-2}$,
$N_A=6.0221415(10)\times 10^{23}\;\mathrm{mol}^{-1}$,
and
$\hbar c = 0.197326968(17)\;\mathrm{GeV}\cdot\mathrm{fm}$ \cite{PDB}.
Therefore, 
\begin{eqnarray}
a & \equiv & 2EV_{CC}
\;=\; (7.6324\times 10^{-5}\,\mathrm{eV}^2)
\times \left( \dfrac{ \rho }{ \mathrm{g/cm}^{3} } \right) 
\times \left( \dfrac{ E }{ \mathrm{GeV} } \right) \;, \cr
b & \equiv & 2EV_{NC}
\;=\; -\frac{1}{2}\,a \;.
\label{aandb}
\end{eqnarray}
For anti-neutrinos, both $V_{CC}$ and $V_{NC}$ reverse their signs.

\section{The Effective Mixing Angles, Neutrino Case}

\subsection{Inclusion of Neutral Current Effects into the Effective Hamiltonian}

The effective potentials derived above enter 
the effective Hamiltonian for neutrino oscillations (multiplied by $2E$) 
as follows:
\begin{equation}
H = 
\tilde{U}
\left[ \begin{array}{ccc} \lambda_1 & 0 & 0 \\
                          0 & \lambda_2 & 0 \\
                          0 & 0 & \lambda_3
       \end{array}
\right]
\tilde{U}^\dagger
= U
\left[ \begin{array}{ccc} 0 & 0 & 0 \\
                          0 & \delta m^2_{21} & 0 \\
                          0 & 0 & \delta m^2_{31}
       \end{array}
\right]
U^\dagger +
\left[ \begin{array}{ccc} a & 0 & 0 \\
                          0 & 0 & 0 \\
                          0 & 0 & 0 
       \end{array}
\right] +
\left[ \begin{array}{ccc} b_e & 0 & 0 \\
                          0 & b_\mu & 0 \\
                          0 & 0 & b_\tau 
       \end{array}
\right] \;.
\label{Hdef}
\end{equation}
Here, $U$ is the MNS matrix in vacuum \cite{MNS},
$a$ comes from the $W$-exchange interaction of $\nu_e$ with the 
electrons in matter, while $b_e$, $b_\mu$, and $b_\tau$ 
come from the $Z$-exchange interaction of each neutrino flavor with the neutrons.
If $b_e = b_\mu = b_\tau = b$, then the $b$-matrix is proportional to the
unit matrix, and it will not contribute to neutrino oscillations.
However, if neutral current universality is broken, then
$b_e\neq b_\mu\neq b_\tau$ in general and the $b$-matrix cannot be ignored.

The experimental constraints from CHARM/CHARM II, Eq.~(\ref{CHARMconstraint}),
allow $b_e$ and $b_\tau$ to deviate significantly from $b=-a/2$, provided that
$b_e+b_\tau = 2b$ to satisfy the $Z$ invisible width constraint, Eq.~(\ref{InvisibleWidthConstraint}).
We therefore write
\begin{equation}
\frac{b_e}{b} \;=\; 1+2\xi \;,\qquad
\frac{b_\mu}{b} \;=\; 1 \;,\qquad
\frac{b_\tau}{b} \;=\; 1-2\xi\;,
\label{xidef}
\end{equation}
and use $\xi$ to parametrize the violation of universality.

Notice that we can rewrite the matter-effect terms in Eq.~(\ref{Hdef})
in several different ways:
\begin{eqnarray}
\lefteqn{\left[ \begin{array}{ccc} a & 0 & 0 \\
                          0 & 0 & 0 \\
                          0 & 0 & 0 
       \end{array}
\right] +
\left[ \begin{array}{ccc} b_e & 0 & 0 \\
                          0 & b_\mu & 0 \\
                          0 & 0 & b_\tau 
       \end{array}
\right]}\cr
& = &
\left[ \begin{array}{ccc} (a+b_e-b_\tau) & 0 & 0 \\
                          0 & (b_\mu-b_\tau) & 0 \\
                          0 & 0 & 0 
       \end{array}
\right] + b_\tau
\left[ \begin{array}{ccc} 1 & 0 & 0 \\
                          0 & 1 & 0 \\
                          0 & 0 & 1 
       \end{array}
\right] \cr
& = &
\left[ \begin{array}{ccc} \left(a+b_e-\dfrac{b_\mu+b_\tau}{2}\right) & 0 & 0 \\
                          0 & \left(\dfrac{b_\mu-b_\tau}{2}\right) & 0 \\
                          0 & 0 & -\left(\dfrac{b_\mu-b_\tau}{2}\right) 
       \end{array}
\right] + \left(\dfrac{b_\mu+b_\tau}{2}\right)
\left[ \begin{array}{ccc} 1 & 0 & 0 \\
                          0 & 1 & 0 \\
                          0 & 0 & 1 
       \end{array}
\right] \;.
\end{eqnarray}
Since the unit matrix terms can be dropped, this shows that
we can always reduce the problem to the case $b_\tau = 0$, or $b_\mu = -b_\tau$.
We will use the latter replacement in the following.
For the case of Eq.~(\ref{xidef}), this entails making the replacement
\begin{equation}
\left[ \begin{array}{ccc} a+b(1+2\xi) & 0 & 0 \\
                          0 & b & 0 \\
                          0 & 0 & b(1-2\xi) 
       \end{array}
\right] 
\Rightarrow
\left[ \begin{array}{ccc} \left(a + 3b\xi \right) & 0 & 0 \\
                          0 & b\xi & 0 \\
                          0 & 0 & -b\xi 
       \end{array}
\right] \;.
\end{equation}
Furthermore, we absorb the factor $3b\xi$ in the $(1,1)$ element 
into $a$ since we can expect $3b\xi\ll a$, and the uncertainty in the matter density $\rho$
which enters into $a$ can be expected to hide any such shift. 
Therefore, the effective Hamiltonian we will consider is
\begin{equation}
H = 
\tilde{U}
\left[ \begin{array}{ccc} \lambda_1 & 0 & 0 \\
                          0 & \lambda_2 & 0 \\
                          0 & 0 & \lambda_3
       \end{array}
\right]
\tilde{U}^\dagger
= U
\left[ \begin{array}{ccc} 0 & 0 & 0 \\
                          0 & \delta m^2_{21} & 0 \\
                          0 & 0 & \delta m^2_{31}
       \end{array}
\right]
U^\dagger +
\left[ \begin{array}{ccc} a & 0 & 0 \\
                          0 & b\xi & 0 \\
                          0 & 0 & -b\xi 
       \end{array}
\right] \;.
\label{Hdef2}
\end{equation}
The problem is to diagonalize $H$ and find the eigenvalues $\lambda_i$ ($i=1,2,3$) and
the diagonalization matrix $\tilde{U}$.

To this end, we use the method of Ref.~\cite{HKOT} in which the $\lambda_i$'s and
$\tilde{U}$ were derived for the $\xi=0$ case.
The procedure followed in Ref.~\cite{HKOT} was to approximately diagonalize 
the effective Hamiltonian, $H$, using the Jacobi method:
$2\times 2$ submatrices of $H$ are diagonalized in the order which
requires the the largest rotation angles until the off-diagonal elements
are negligibly small.
As the order parameter to evaluate the size of these off-diagonal elements,
we use
\begin{equation}
\varepsilon \equiv \sqrt{\dfrac{\delta m^2_{21}}{|\delta m^2_{31}|}}\;,
\end{equation}
and consider $H$ to be approximately diagonalized when the rotation angles required for
further diagonalization are of order $\varepsilon^3$ or smaller.
For $\delta m^2_{21} = 8.2^{+0.6}_{-0.5}\times 10^{-5}\,\mathrm{eV}^2$ and
$|\delta m^2_{31}| = (1.5\sim 3.4)\times 10^{-3}\,\mathrm{eV}^2$, 
we have $\varepsilon = 0.15\sim 0.24$ and $\varepsilon^3 = 0.0034\sim 0.014$.

For the sizes of the mixing angles in vacuum, we assume $\theta_{13} = O(\varepsilon)$,
$\cos(2\theta_{12})/2 = O(\varepsilon)$, and
$\cos(2\theta_{23}) \le O(\varepsilon)$ as in Ref.~\cite{HKOT}.
We also assume that the universality violation parameter
$\xi$ is of order $\varepsilon^2 = 0.02\sim 0.06$, since the
central value of the CHARM/CHARM II result translates to $\xi = 0.025$.

\subsection{Diagonalization of the Effective Hamiltonian}

For the neutral current term $b\xi$ in Eq.~(\ref{Hdef2}) 
to have a non-negligible effect on neutrino oscillations, 
we anticipate that it must be at least as large as, or larger than, the smaller 
mass-squared-difference $\delta m^2_{21}$.
Since we have assumed $\xi = O(\varepsilon^2)$,  
this requires $a = -2b = 2\sqrt{2}G_F N E$ to be at least as large as,
or larger than, the larger mass-squared-difference $|\delta m^2_{31}|$.
For the sake of concreteness, we will consider the case 
$a/|\delta m^2_{31}| = O(\varepsilon^{-1})$ in the following.
%

Introducing the matrix
\begin{equation}
\mathcal{Q} = \mathrm{diag}(1,1,e^{i\delta})\;,
\label{Qdef}
\end{equation}
we begin by partially diagonalizing the Hamiltonian $H$ as
\begin{eqnarray}
H' & = & \mathcal{Q}^\dagger U^\dagger H U \mathcal{Q}\cr
& = &
\left[ \begin{array}{ccc} 0 & 0 & 0 \\
                          0 & \delta m^2_{21} & 0 \\
                          0 & 0 & \delta m^2_{31}
       \end{array}
\right] +
\mathcal{Q}^\dagger U^\dagger
\left[ \begin{array}{ccc} a & 0 & 0 \\
                          0 & b\xi & 0 \\
                          0 & 0 & -b\xi 
       \end{array}
\right] 
U \mathcal{Q}  \cr
& = &
\left[ \begin{array}{ccc} 0 & 0 & 0 \\
                          0 & \delta m^2_{21} & 0 \\
                          0 & 0 & \delta m^2_{31}
       \end{array}
\right] + a\,\mathcal{Q}^\dagger
\left[ \begin{array}{ccc} U^*_{e1}U_{e1} & U^*_{e1}U_{e2} & U^*_{e1}U_{e3} \\
                          U^*_{e2}U_{e1} & U^*_{e2}U_{e2} & U^*_{e2}U_{e3} \\
                          U^*_{e3}U_{e1} & U^*_{e3}U_{e2} & U^*_{e3}U_{e3} 
       \end{array}
\right] \mathcal{Q} \cr
& &  + b\xi\,\mathcal{Q}^\dagger \left\{
\left[ \begin{array}{ccc} U^*_{\mu 1}U_{\mu 1} & U^*_{\mu 1}U_{\mu 2} & U^*_{\mu 1}U_{\mu 3} \\
                          U^*_{\mu 2}U_{\mu 1} & U^*_{\mu 2}U_{\mu 2} & U^*_{\mu 2}U_{\mu 3} \\
                          U^*_{\mu 3}U_{\mu 1} & U^*_{\mu 3}U_{\mu 2} & U^*_{\mu 3}U_{\mu 3} 
       \end{array}
\right] - 
\left[ \begin{array}{ccc} U^*_{\tau 1}U_{\tau 1} & U^*_{\tau 1}U_{\tau 2} & U^*_{\tau 1}U_{\tau 3} \\
                          U^*_{\tau 2}U_{\tau 1} & U^*_{\tau 2}U_{\tau 2} & U^*_{\tau 2}U_{\tau 3} \\
                          U^*_{\tau 3}U_{\tau 1} & U^*_{\tau 3}U_{\tau 2} & U^*_{\tau 3}U_{\tau 3} 
       \end{array}
\right] \right\} \mathcal{Q}\;.
\end{eqnarray}
The matrix multiplying $a$ is given by
\begin{equation}
M_a \;=\; 
\mathcal{Q}^\dagger
\left[ \begin{array}{ccc} U^*_{e1}U_{e1} & U^*_{e1}U_{e2} & U^*_{e1}U_{e3} \\
                          U^*_{e2}U_{e1} & U^*_{e2}U_{e2} & U^*_{e2}U_{e3} \\
                          U^*_{e3}U_{e1} & U^*_{e3}U_{e2} & U^*_{e3}U_{e3} 
       \end{array}
\right] \mathcal{Q}
\;=\;
\left[ \begin{array}{ccc}
       c_{12}^2 c_{13}^2    & c_{12}s_{12}c_{13}^2 & c_{12}c_{13}s_{13} \\
       c_{12}s_{12}c_{13}^2 & s_{12}^2 c_{13}^2    & s_{12}c_{13}s_{13} \\
       c_{12}c_{13}s_{13}   & s_{12}c_{13}s_{13}   & s_{13}^2
       \end{array}
\right]\;.
\end{equation}
Using $\theta_{13} = O(\varepsilon)$, we estimate the sizes of the elements of $M_a$ to be
\begin{equation}
M_a \;=\;
\left[ \begin{array}{ccc}
       O(1) & O(1) & O(\varepsilon) \\
       O(1) & O(1) & O(\varepsilon) \\
       O(\varepsilon) & O(\varepsilon) & O(\varepsilon^2)
       \end{array}
\right]\;.
\end{equation}
The matrix multiplying $b\xi$ is given by
\begin{eqnarray}
M_b & = &
\mathcal{Q}^\dagger\left\{
\left[ \begin{array}{ccc} U^*_{\mu 1}U_{\mu 1} & U^*_{\mu 1}U_{\mu 2} & U^*_{\mu 1}U_{\mu 3} \\
                          U^*_{\mu 2}U_{\mu 1} & U^*_{\mu 2}U_{\mu 2} & U^*_{\mu 2}U_{\mu 3} \\
                          U^*_{\mu 3}U_{\mu 1} & U^*_{\mu 3}U_{\mu 2} & U^*_{\mu 3}U_{\mu 3} 
       \end{array}
\right] - 
\left[ \begin{array}{ccc} U^*_{\tau 1}U_{\tau 1} & U^*_{\tau 1}U_{\tau 2} & U^*_{\tau 1}U_{\tau 3} \\
                          U^*_{\tau 2}U_{\tau 1} & U^*_{\tau 2}U_{\tau 2} & U^*_{\tau 2}U_{\tau 3} \\
                          U^*_{\tau 3}U_{\tau 1} & U^*_{\tau 3}U_{\tau 2} & U^*_{\tau 3}U_{\tau 3} 
       \end{array}
\right] \right\} \mathcal{Q} \cr
& = &
\left[ \begin{array}{l}
\phantom{-}(s_{12}^2-c_{12}^2 s_{13}^2)\cos(2\theta_{23}) 
+ \sin(2\theta_{12})\sin(2\theta_{23})s_{13}\cos\delta \\
-(1+s_{13}^2)s_{12}c_{12}\cos(2\theta_{23}) 
-(c_{12}^2 e^{-i\delta} - s_{12}^2 e^{+i\delta})s_{13}\sin(2\theta_{23}) \\
-s_{12}c_{13}\sin(2\theta_{23})e^{-i\delta} + c_{12}s_{13}c_{13}\cos(2\theta_{23}) 
\end{array} \right. \cr
&   & \qquad\qquad
\begin{array}{l}
-(1+s_{13}^2)s_{12}c_{12}\cos(2\theta_{23}) 
-(c_{12}^2 e^{+i\delta} - s_{12}^2 e^{-i\delta})s_{13}\sin(2\theta_{23}) \\
\phantom{-}(c_{12}^2-s_{12}^2 s_{13}^2)\cos(2\theta_{23}) 
- \sin(2\theta_{12})\sin(2\theta_{23})s_{13}\cos\delta \\
\phantom{-}c_{12}c_{13}\sin(2\theta_{23})e^{-i\delta} + s_{12}s_{13}c_{13}\cos(2\theta_{23}) 
\end{array} \cr
&   & \qquad\qquad\qquad\qquad
\left. \begin{array}{l}
-s_{12}c_{13}\sin(2\theta_{23})e^{+i\delta} + c_{12}s_{13}c_{13}\cos(2\theta_{23}) \\
\phantom{-}c_{12}c_{13}\sin(2\theta_{23})e^{+i\delta} + s_{12}s_{13}c_{13}\cos(2\theta_{23}) \\
 -c_{13}^2\cos(2\theta_{23})
\end{array} \right] \;.
\end{eqnarray}
Using $\cos(2\theta_{23}) \le O(\varepsilon)$ and
$\theta_{13} = O(\varepsilon)$, we estimate the sizes of the elements of $M_b$
to be
\begin{equation}
M_b = \left[ \begin{array}{ccc} O(\varepsilon) & O(\varepsilon) & O(1) \\
                                O(\varepsilon) & O(\varepsilon) & O(1) \\
                                O(1)           & O(1)           & O(\varepsilon)
             \end{array}
      \right]\;.
\end{equation}
%
%
Since we are only interested in the leading order effect in $\xi$,
we neglect the $O(\varepsilon)$ terms in $M_b$ and approximate
\begin{equation}
M_b \approx
\left[ \begin{array}{ccc} 0                  & 0                  & -s_{12}e^{+i\delta} \\
                          0                  & 0                  & \phantom{-}c_{12}e^{+i\delta} \\
                         -s_{12}e^{-i\delta} & c_{12}e^{-i\delta} &  0
       \end{array}
\right]\;.
\end{equation}
Under this approximation, the effective Hamiltonian that must be diagonalized is
\begin{eqnarray}
H' 
& = & \mathrm{diag}(0,\delta m^2_{21},\delta m^2_{31}) + a\,M_a + b\xi\,M_b \cr
& = & 
\left[ 
\begin{array}{ccc}
a\,c_{12}^2 c_{13}^2    & a\,c_{12}s_{12}c_{13}^2 & a\,c_{12}c_{13}s_{13}-b\xi\,s_{12}e^{+i\delta} \\
a\,c_{12}s_{12}c_{13}^2 & a\,s_{12}^2 c_{13}^2 + \delta m^2_{21}   & a\,s_{12}c_{13}s_{13}+b\xi\,c_{12}e^{+i\delta} \\
a\,c_{12}c_{13}s_{13}-b\xi\,s_{12}e^{-i\delta}   & a\,s_{12}c_{13}s_{13}+b\xi\,c_{12}e^{-i\delta} & a\,s_{13}^2 + \delta m^2_{31}
\end{array}
\right]\;.
\end{eqnarray}
At this point, we set $\delta=0$ for the sake of simplicity.
Then $H'$ becomes
\begin{equation}
H' =
\left[ 
\begin{array}{ccc}
a\,c_{12}^2 c_{13}^2            & a\,c_{12}s_{12}c_{13}^2                & a\,c_{12}c_{13}s_{13}-b\xi\,s_{12} \\
a\,c_{12}s_{12}c_{13}^2         & a\,s_{12}^2 c_{13}^2 + \delta m^2_{21} & a\,s_{12}c_{13}s_{13}+b\xi\,c_{12} \\
a\,c_{12}c_{13}s_{13}-b\xi\,s_{12} & a\,s_{12}c_{13}s_{13}+b\xi\,c_{12}        & a\,s_{13}^2 + \delta m^2_{31}
\end{array}
\right] 
= a
\left[ \begin{array}{ccc}
       O(1) & O(1) & O(\varepsilon) \\
       O(1) & O(1) & O(\varepsilon) \\
       O(\varepsilon) & O(\varepsilon) & O(\varepsilon)
       \end{array}
\right]
\;.
\end{equation}
%

\subsubsection{First Rotation}

Applying the Jacobi method to $H'$,
we first diagonalize the $(1,2)$ submatrix which requires a rotation
by an angle of $O(1)$. 
Define the matrix $V$ as:
\begin{equation}
V = 
\left[ \begin{array}{ccc} c_{\varphi} &  s_{\varphi} & 0 \\
	                     -s_{\varphi} &  c_{\varphi} & 0 \\
	                      0 & 0 & 1
	   \end{array}
\right]\;,
\label{Vdef}
\end{equation}
where
\begin{equation}
c_{\varphi} =\cos\varphi\;,\quad
s_{\varphi} =\sin\varphi\;,\quad
\tan 2\varphi \equiv 
\dfrac{a c_{13}^2\sin2\theta_{12}}{\delta m^2_{21}-a c_{13}^2\cos2\theta_{12}}\;,\quad
\left(0\le\varphi<\frac{\pi}{2}-\theta_{12}\right)\;.
\label{phi1def}
\end{equation}
Then,
\begin{equation}
H'' 
= V^\dagger H' V 
=
\left[ \begin{array}{ccc}
       \lambda'_1 & 0 & a {c}_{12}' c_{13}s_{13} - b\xi s_{12}' \\
       0 & \lambda'_2 & a {s}_{12}' c_{13}s_{13} + b\xi c_{12}' \\
       a {c}_{12}' c_{13}s_{13} - b\xi s_{12}' & a {s}_{12}' c_{13}s_{13} + b\xi c_{12}' &
       a s_{13}^2 + \delta m^2_{31}
       \end{array}
\right] \;,
\label{Hdoubleprimedef}
\end{equation}
where
\begin{equation}
{c}_{12}' = \cos\theta_{12}' \;,\quad
{s}_{12}' = \sin\theta_{12}' \;,\quad
{\theta}_{12}' = \theta_{12} + \varphi \;,
\label{theta12primedef}
\end{equation}
and
\begin{eqnarray}
\lambda'_1 
& = & \dfrac{(a c_{12}^2 c_{13}^2) c^2_\varphi
           - (a s_{12}^2 c_{13}^2 + \delta m^2_{21}) s^2_\varphi}
            {c^2_\varphi - s^2_\varphi} 
\;=\; \lambda'_{-} \;,\cr
\lambda'_2 
& = & \dfrac{(a s_{12}^2 c_{13}^2 + \delta m^2_{21}) c^2_\varphi
           - (a c_{12}^2 c_{13}^2) s^2_\varphi}
            {c^2_\varphi - s^2_\varphi}
\;=\; \lambda'_{+} \;,
\label{lambdaprimesdef}
\end{eqnarray}
with
\begin{equation}
\lambda'_{\pm}
= \dfrac{ (a c_{13}^2+\delta m^2_{21})
          \pm\sqrt{ (a c_{13}^2-\delta m^2_{21})^2 + 4 a c_{13}^2 s_{12}^2 \delta m^2_{21} }
        }
        { 2 }\;.
\label{lambdaprimeplusminusdef}
\end{equation}
As discussed in Ref.~\cite{HKOT}, in the region $a/|\delta m^2_{31}|=O(\varepsilon^{-1})$,
we can expand $\theta'_{12}$ as
\begin{equation}
\theta'_{12}
= \frac{\pi}{2} - \frac{\delta m^2_{21}}{2a}\sin(2\theta_{12}) + O(\varepsilon^5) \;,
\label{thetaprime12}
\end{equation}
from which we can conclude 
\begin{eqnarray}
s'_{12} 
& \approx & \sin\left(\frac{\pi}{2}-\frac{\delta m^2_{21}}{2a}\sin(2\theta_{12}) \right)
\;=\; \cos\left(\frac{\delta m^2_{21}}{2a}\sin(2\theta_{12}) \right)
\;=\; 1 - O(\varepsilon^{6}) \;, \cr 
c'_{12} 
& \approx & \cos\left(\frac{\pi}{2}-\frac{\delta m^2_{21}}{2a}\sin(2\theta_{12}) \right)
\;=\; \sin\left(\frac{\delta m^2_{21}}{2a}\sin(2\theta_{12}) \right)
\;=\; O(\varepsilon^3) \;.
\end{eqnarray}
Also, expanding $\lambda'_\pm$, we find
\begin{eqnarray}
\lambda'_- 
& = & \delta m^2_{21} c^2_{12} + a\,O(\varepsilon^6)
\;=\; a\,O(\varepsilon^3) \;,\cr
\lambda'_+
& = & a c_{13}^2 + \delta m^2_{21} s^2_{12} + a\,O(\varepsilon^6)
\;=\; a\,O(1) \;.
\end{eqnarray}
Therefore, the sizes of the elements of $H''$ are evaluated to be
\begin{equation}
H'' = a\,
\left[ \begin{array}{ccc}
       O(\varepsilon^3) & 0                & O(\varepsilon^2) \\
       0                & O(1)             & O(\varepsilon)   \\
       O(\varepsilon^2) & O(\varepsilon)   & O(\varepsilon)
       \end{array}
\right] \;.
\end{equation}
Unlike the $\xi=0$ case considered in Ref.~\cite{HKOT}, 
both the $(1,3)$ and $(2,3)$ submatrices require rotations by angles of
$O(\varepsilon)$ to diagonalize.
Here, we diagonalize the $(2,3)$ submatrix next to maintain the parallel with
the $\xi=0$ case.

\subsubsection{Second Rotation}

The matrix $W$ which diagonalizes the $(2,3)$ submatrix is
\begin{equation}
W = 
\left[ \begin{array}{ccc} 1 & 0 & 0 \\
                          0 &  c_{\phi} &  s_{\phi} \\
	                      0 & -s_{\phi} &  c_{\phi} 
	   \end{array}
\right]\;,
\label{Wdef}
\end{equation}
where $c_{\phi}=\cos\phi$, $s_{\phi}=\sin\phi$, and
\begin{eqnarray}
\tan 2\phi & \equiv & 
\dfrac{2(a s'_{12}s_{13}c_{13} + b\xi c'_{12})}
      {\delta m^2_{31}+a s_{13}^2 - \lambda'_2} \cr
& = &
\dfrac{2a s'_{12}s_{13}c_{13}}
      {\delta m^2_{31}+a s_{13}^2 - \lambda'_2}
\left\{ 1 - \xi\left(\dfrac{c'_{12}}{2 s'_{12}s_{13}c_{13}}\right) \right\} \cr
& = & \tan 2\phi_0
\left\{ 1 - \xi\left(\dfrac{c'_{12}}{2 s'_{12}s_{13}c_{13}}\right) \right\} 
\;.
\label{phi2def}
\end{eqnarray}
The angle $\phi$ is in the first quadrant when $\delta m^2_{31} > 0$ (normal hierarchy),
and in the fourth quadrant when $\delta m^2_{31} < 0$ (inverted hierarchy).
$\phi_0$ is the rotation angle when $\xi=0$.
Taking the arc-tangent of both sides of the above equation, we find
\begin{equation}
\phi = \phi_0
- \xi \left(\dfrac{\sin 4\phi_0\,\cot\theta'_{12}}{4\sin 2\theta_{13}}\right)
+ O(\xi^2) \;.
\end{equation}
When $a/|\delta m^2_{31}| = O(\varepsilon^{-1})$, 
$\phi_0$ is given by \cite{HKOT}
\begin{equation}
\phi_0 = 
\left\{ \begin{array}{ll}
        \left( \dfrac{\pi}{2} - \theta_{13} \right)
        - \dfrac{\delta m^2_{31}}{a}\,\theta_{13}+ O(\varepsilon^3) &
        \qquad(\delta m^2_{31}>0) \;, \\
        -\theta_{13} - \dfrac{\delta m^2_{31}}{a}\,\theta_{13} + O(\varepsilon^3) &
        \qquad(\delta m^2_{31}<0) \;.
        \end{array}
\right. 
\label{phizero}
\end{equation}
Therefore, $\sin 4\phi_0 \approx -\sin(4\theta_{13})$ for both the 
$\delta m^2_{31} >0$ and $\delta m^2_{31} <0$ cases and
using Eq.~(\ref{thetaprime12}), we find
\begin{equation}
\phi \approx 
\phi_0
+ \xi \left(\dfrac{\delta m^2_{21}}{4a}\right)\sin(2\theta_{12}) 
= \phi_0 + \frac{1}{4}O(\varepsilon^5)\;.
\end{equation}
Therefore, the difference between $\phi$ and $\phi_0$ can be neglected
in this range.

Using $W$, we obtain
\begin{eqnarray}
H'''
& = & W^\dagger H'' W \cr
& = &
\left[ \begin{array}{ccc}
\lambda'_1 
& -(a c'_{12}c_{13}s_{13}-b\xi s_{12}' )s_{\phi} 
&  (a c'_{12}c_{13}s_{13}-b\xi s_{12}' )c_{\phi} \\
-(a c'_{12}c_{13}s_{13}-b\xi s_{12}' )s_{\phi} & \lambda''_2 & 0  \\
 (a c'_{12}c_{13}s_{13}-b\xi s_{12}' )c_{\phi} & 0 & \lambda''_3
\end{array} \right] \;,\cr
& &
\label{Htripleprimedef}
\end{eqnarray}
where
\begin{eqnarray}
\lambda''_2 & = &  
\dfrac{\lambda'_2 c^2_\phi - (a s^2_{13}+\delta m^2_{31})s^2_\phi}{c^2_\phi-s^2_\phi} \;,\cr 
\lambda''_3 & = &  
\dfrac{(a s^2_{13}+\delta m^2_{31})c^2_\phi - \lambda'_2 s^2_\phi}{c^2_\phi-s^2_\phi} \;.
\label{lambdadoubleprimes}
\end{eqnarray}
If we define
\begin{equation}
\lambda''_{\pm} \equiv
   \dfrac{ [ \lambda'_{2} + (a s_{13}^2+\delta m^2_{31}) ]
\pm \sqrt{ [ \lambda'_{2} - (a s_{13}^2+\delta m^2_{31}) ]^2 
         + 4 ( a {s'_{12}}c_{13}s_{13} + b\xi c_{12}' )^2 }
         }
         { 2 } \;,
\label{lambdadoubleprimeplusminusdef}
\end{equation}
then
\begin{eqnarray}
\lambda''_2 = \lambda''_{-}\;,\quad
\lambda''_3 = \lambda''_{+}\;,\qquad \mbox{if $\delta m^2_{31} > 0$}\;, \cr
\lambda''_2 = \lambda''_{+}\;,\quad
\lambda''_3 = \lambda''_{-}\;,\qquad \mbox{if $\delta m^2_{31} < 0$}\;.
\end{eqnarray}
When $a/|\delta m^2_{31}| = O(\varepsilon^{-1})$, we can expand 
$\lambda''_\pm$ as
\begin{eqnarray}
\lambda''_{-} 
& = & \delta m^2_{31} c_{13}^2 + O(\varepsilon^3 |\delta m^2_{31}|)
\;=\; a\,O(\varepsilon) \;, \cr
\lambda''_{+}
& = & a + \delta m^2_{31} s_{13}^2 + \delta m^2_{21} s_{12}^2
        + O(\varepsilon^3 |\delta m^2_{31}|)
\;=\; a\,O(1) \;.
\end{eqnarray}
Also, from Eq.~(\ref{phizero}) and the fact that $\phi\approx\phi_0$, 
we conclude
\begin{equation}
\begin{array}{lll}
s_\phi \approx c_{13} = O(1)\;,\quad & 
c_\phi \approx s_{13} = O(\varepsilon)\;,\quad &
(\delta m^2_{31} > 0)\;, \\
s_\phi \approx -s_{13} = O(\varepsilon)\;,\quad & 
c_\phi \approx c_{13} = O(1)\;,\quad &
(\delta m^2_{31} < 0)\;.
\end{array}
\end{equation}
In the $\xi=0$ case considered in Ref.~\cite{HKOT},
$H'''$ was already approximately diagonal and further diagonalization was not necessary.
However, when $\xi=O(\varepsilon^2)$, 
the sizes of the elements of $H'''$ are found to be
\begin{equation}
H''' = a
\left[ \begin{array}{ccc}
       O(\varepsilon^3) & O(\varepsilon^2) & O(\varepsilon^3) \\
       O(\varepsilon^2) & O(\varepsilon)   & 0 \\
       O(\varepsilon^3) & 0                & O(1) 
       \end{array}
\right] 
\end{equation}
when $\delta m^2_{31}>0$, and
\begin{equation}
H''' = a
\left[ \begin{array}{ccc}
       O(\varepsilon^3) & O(\varepsilon^3) & O(\varepsilon^2) \\
       O(\varepsilon^3) & O(1)             & 0 \\
       O(\varepsilon^2) & 0                & O(\varepsilon) 
       \end{array}
\right] 
\end{equation}
when $\delta m^2_{31}<0$.
Therefore, for the $\delta m^2_{31}>0$ case (normal hierarchy)
we must diagonalize the $(1,2)$ submatrix next, while for the 
$\delta m^2_{31}<0$ case (inverted hierarchy) 
we must diagonalize the $(1,3)$ submatrix next.

\subsubsection{Third Rotation, $\delta m^2_{31}>0$ Case}

Define the matrix $X$ as
\begin{equation}
X = 
\left[ \begin{array}{ccc} c_\chi & s_\chi & 0 \\
                         -s_\chi & c_\chi & 0 \\
                          0      & 0      & 1
       \end{array}
\right] \;,
\end{equation}
where
\begin{equation}
c_\chi = \cos\chi\;,\quad
s_\chi = \sin\chi\;,\quad
\tan 2\chi \equiv \frac{ -2(a c'_{12}c_{13}s_{13}- b\xi s'_{12})s_\phi }
                       { \lambda''_{2} - \lambda'_{1} }\;.
\label{chidef}
\end{equation}
Then,
\begin{eqnarray}
H''''_X 
& = & X^\dagger H''' X \cr
& = & 
\left[ \begin{array}{ccc}
\lambda'''_{1X} & 0               & (a c'_{12}c_{13}s_{13} - b\xi s'_{12})c_\phi c_\chi \\
0               & \lambda'''_{2X} & (a c'_{12}c_{13}s_{13} - b\xi s'_{12})c_\phi s_\chi \\
(a c'_{12}c_{13}s_{13} - b\xi s'_{12})c_\phi c_\chi &
(a c'_{12}c_{13}s_{13} - b\xi s'_{12})c_\phi s_\chi &
\lambda''_3
\end{array} \right]\;, \cr
& &
\end{eqnarray}
where
\begin{eqnarray}
\lambda'''_{1X} & = &  
\dfrac{\lambda'_1 c^2_\chi - \lambda''_2 s^2_\chi}{c^2_\chi-s^2_\chi} 
\;=\; \lambda'''_{X-}\;,\cr 
\lambda'''_{2X} & = &  
\dfrac{\lambda''_2 c^2_\chi - \lambda'_1 s^2_\chi}{c^2_\chi-s^2_\chi} 
\;=\; \lambda'''_{X+}\;,
\label{lambdatripleprimes}
\end{eqnarray}
with
\begin{equation}
\lambda'''_{X\pm} \equiv
   \dfrac{ ( \lambda''_{2} + \lambda'_{1} )
\pm \sqrt{ ( \lambda''_{2} - \lambda'_{1} )^2 
         + 4 ( a {c'_{12}}c_{13}s_{13} - b\xi s_{12}' )^2 s_\phi^2 }
         }
         { 2 } \;.
\label{lambdatripleprimeplusminusdef}
\end{equation}
Recalling that
\begin{eqnarray}
a c'_{12}c_{13}s_{13} 
& \approx & a\left(\dfrac{\delta m^2_{21}}{2a}\sin(2\theta_{12})\right) c_{13}s_{13}
\;=\; a\,O(\varepsilon^4) \;,\cr
b\xi s'_{12} 
& \approx & -\frac{a}{2}\xi
\;=\; -a\,O(\varepsilon^2) \;, \cr
\lambda'_1 
& \approx & \delta m^2_{21} c_{12}^2
\;=\; a\,O(\varepsilon^3) \;, \cr
\lambda''_2 
& \approx & \delta m^2_{31} c_{13}^2
\;=\; a\,O(\varepsilon) \;,
\end{eqnarray}
and $s_\phi \approx 1$, we find 
\begin{equation}
\tan 2\chi =
\dfrac{-2(a c'_{12}c_{13}s_{13}-b\xi s'_{12})s_\phi}{\lambda''_2 - \lambda'_1}
= \dfrac{2b\xi}{\delta m^2_{31}}\left\{ 1 + O(\varepsilon^2) \right\} \;.
\end{equation}
Therefore, the angle $\chi$ is given approximately by 
\begin{equation}
\chi \approx \frac{b\xi}{\delta m^2_{31}} = O(\varepsilon)\;,
\end{equation}
from which we can conclude that $s_\chi = O(\varepsilon)$ and $c_\chi = O(1)$.
The eigenvalues can also be expanded in $\varepsilon$ and we find
\begin{eqnarray}
\lambda'''_{1X}
& = & \delta m^2_{21} c_{12}^2 - \dfrac{b^2\xi^2}{\delta m^2_{31}} + a\,O(\varepsilon^5)
\;=\; a\,O(\varepsilon^3) \;,\cr
\lambda'''_{2X}
& = & \delta m^2_{31} c_{13}^2 + \dfrac{b^2\xi^2}{\delta m^2_{31}} + a\,O(\varepsilon^4)
\;=\; a\,O(\varepsilon) \;.
\end{eqnarray}
Note that these shifts of the eigenvalues are of order $a O(\varepsilon^3)$ and have 
negligible effect on $\delta\lambda_{31} = a O(1)$ or $\delta\lambda_{21} = a O(\varepsilon)$.
Putting everything together, we evaluate the sizes of the elements of $H''''_X$ to find
\begin{equation}
H''''_X = a
\left[ \begin{array}{ccc}
       O(\varepsilon^3) & 0                & O(\varepsilon^3) \\
       0                & O(\varepsilon)   & O(\varepsilon^4) \\
       O(\varepsilon^3) & O(\varepsilon^4) & O(1)
       \end{array}
\right]\;.
\end{equation}
This shows that
further diagonalization requires rotations by angles of $O(\varepsilon^3)$ or
smaller, which we will neglect.

Thus, we have found that when $\xi=O(\varepsilon^2)$ and $\delta m^2_{31}>0$
(normal hierarchy),
we need an extra $(1,2)$-rotation to diagonalize $H$, 
and the diagonalization matrix is $UVWX$, which we need to identify with
\begin{equation}
\tilde{U} =
\left[ \begin{array}{ccc} 1 & 0 & 0 \\
                          0 &  \tilde{c}_{23} & \tilde{s}_{23} \\
                          0 & -\tilde{s}_{23} & \tilde{c}_{23}
       \end{array}
\right]
\left[ \begin{array}{ccc} \tilde{c}_{13} & 0 & \tilde{s}_{13} e^{-i\tilde{\delta}} \\
                          0 & 1 & 0 \\
                          -\tilde{s}_{13} e^{i\tilde{\delta}} & 0 & \tilde{c}_{13}
       \end{array}
\right]
\left[ \begin{array}{ccc} \tilde{c}_{12} & \tilde{s}_{12} & 0 \\
                         -\tilde{s}_{12} & \tilde{c}_{12} & 0 \\
                          0 & 0 & 1
       \end{array}
\right]  
\end{equation}
to obtain the effective mixing angles and effective CP phase.
From Ref.~\cite{HKOT}, we know that when
$\delta m^2_{31}>0$ and $a/|\delta m^2_{31}| = O(\varepsilon^{-1})$,
identification of $\tilde{U}$ with $UVW$ leads to
\begin{eqnarray}
\tilde{\theta}_{13} & = & \theta'_{13} \;,\cr
\tilde{\theta}_{12} & = & \frac{\pi}{2} - \frac{c_{13}}{c'_{13}}
\left(\frac{\delta m^2_{21}}{2a}\right)\sin(2\theta_{12}) \;,\cr
\tilde{\theta}_{23} & = & \theta_{23} + \frac{s_\phi}{c'_{13}}
\left(\frac{\delta m^2_{21}}{2a}\right)\sin(2\theta_{12}) \;,\cr
\tilde{\delta} & = & 0 \;,
\end{eqnarray}
where we have defined
\begin{equation}
\theta'_{13}\equiv \theta_{13}+\phi\;,\qquad
c'_{13} = \cos\theta'_{13}\;,\qquad
s'_{13} = \sin\theta'_{13}\;.
\end{equation}
(Recall that we are considering the $\delta=0$ case).
Since $X$ is an $(1,2)$-rotation, multiplication of $\tilde{U}=UVW$ 
from the right by $X$ only shifts the value of $\tilde{\theta}_{12}$ by $\chi$.
Therefore, we can conclude that
\begin{eqnarray}
\tilde{\theta}_{13} & = & \theta'_{13} \;,\cr
\tilde{\theta}_{12} & = & \frac{\pi}{2} - \frac{c_{13}}{c'_{13}}
\left(\frac{\delta m^2_{21}}{2a}\right)\sin(2\theta_{12}) + \chi \;,\cr
\tilde{\theta}_{23} & = & \theta_{23} + \frac{s_\phi}{c'_{13}}
\left(\frac{\delta m^2_{21}}{2a}\right)\sin(2\theta_{12}) \;, \cr
\tilde{\delta} & = & 0\;.
\end{eqnarray}
In these expressions, non only $\chi$, but also $\phi$ and 
$\theta'_{13} = \theta_{13} + \phi$ depend on $\xi$.
However, the $\xi$-dependence of $\phi$ is very weak.
The $\xi$-dependence of $\delta\lambda_{31}$ and $\delta\lambda_{21}$ are also
weak, so the effect of a non-zero $\xi$ will appear dominantly in
$\tilde{\theta}_{12}$.

\subsubsection{Third Rotation, $\delta m^2_{31}<0$ Case}

In this case, we need to diagonalize the $(1,3)$-submatrix of $H'''$.
Define the matrix $Y$ as
\begin{equation}
Y = 
\left[ \begin{array}{ccc} c_\eta & 0 & s_\eta \\
                          0      & 1 & 0      \\
                         -s_\eta & 0 & c_\eta
       \end{array}
\right] \;,
\end{equation}
where
\begin{equation}
c_\eta = \cos\eta\;,\quad
s_\eta = \sin\eta\;,\quad
\tan 2\eta \equiv \frac{ 2(a c'_{12}c_{13}s_{13}- b\xi s'_{12})c_\phi }
                       { \lambda''_{3} - \lambda'_{1} }\;.
\end{equation}
Then,
\begin{eqnarray}
H''''_Y 
& = & Y^\dagger H''' Y \cr
& = & 
\left[ \begin{array}{ccc}
\lambda'''_{1Y} & -(a c'_{12}c_{13}s_{13} - b\xi s'_{12})s_\phi c_\eta & 0 \\
-(a c'_{12}c_{13}s_{13} - b\xi s'_{12})s_\phi c_\eta & 
\lambda''_2 & 
-(a c'_{12}c_{13}s_{13} - b\xi s'_{12})s_\phi s_\eta \\
0 & -(a c'_{12}c_{13}s_{13} - b\xi s'_{12})s_\phi s_\eta &
\lambda'''_{3Y}
\end{array} \right]\;, \cr
& &
\end{eqnarray}
where,
\begin{eqnarray}
\lambda'''_{1Y} & = &  
\dfrac{\lambda'_1 c^2_\eta - \lambda''_3 s^2_\eta}{c^2_\eta-s^2_\eta} 
\;=\; \lambda'''_{Y+}\;,\cr 
\lambda'''_{3Y} & = &  
\dfrac{\lambda''_3 c^2_\eta - \lambda'_1 s^2_\eta}{c^2_\eta-s^2_\eta} 
\;=\; \lambda'''_{Y-}\;,
\label{lambdaquadrupleprimes}
\end{eqnarray}
with
\begin{equation}
\lambda'''_{Y\pm} \equiv
   \dfrac{ ( \lambda''_{3} + \lambda'_{1} )
\pm \sqrt{ ( \lambda''_{3} - \lambda'_{1} )^2 
         + 4 ( a {c'_{12}}c_{13}s_{13} + b\xi s_{12}' )^2 c_\phi^2 }
         }
         { 2 } \;.
\label{lambdaquadrupleprimeplusminusdef}
\end{equation}
Using
\begin{eqnarray}
a c'_{12}c_{13}s_{13} 
& \approx & a\left(\dfrac{\delta m^2_{21}}{2a}\sin(2\theta_{12})\right) c_{13}s_{13}
\;=\; a\,O(\varepsilon^4) \;,\cr
b\xi s'_{12} 
& \approx & -\frac{a}{2}\xi
\;=\; -a\,O(\varepsilon^2) \;, \cr
\lambda'_1 
& \approx & \delta m^2_{21} c_{12}^2
\;=\; a\,O(\varepsilon^3) \;, \cr
\lambda''_3 
& \approx & \delta m^2_{31} c_{13}^2
\;=\; a\,O(\varepsilon) \;,
\end{eqnarray}
and $c_\phi \approx 1$, we find 
\begin{equation}
\tan 2\eta =
\dfrac{2(a c'_{12}c_{13}s_{13}-b\xi s'_{12})c_\phi}{\lambda''_3 - \lambda'_1}
= -\dfrac{2b\xi}{\delta m^2_{31}}\left\{ 1 + O(\varepsilon^2) \right\} \;.
\end{equation}
Therefore, the angle $\eta$ is given approximately by 
\begin{equation}
\eta \approx -\frac{b\xi}{\delta m^2_{31}} 
= \frac{b\xi}{|\delta m^2_{31}|} = O(\varepsilon)\;,
\end{equation}
from which we can conclude that $s_\eta = O(\varepsilon)$ and $c_\eta = O(1)$.
The eigenvalues can also be expanded in $\varepsilon$ and we find
\begin{eqnarray}
\lambda'''_{1Y}
& = & \delta m^2_{21} c_{12}^2 - \dfrac{b^2\xi^2}{\delta m^2_{31}} + a\,O(\varepsilon^5)
\;=\; a\,O(\varepsilon^3)\;,\cr
\lambda'''_{3Y}
& = & \delta m^2_{31} c_{13}^2 + \dfrac{b^2\xi^2}{\delta m^2_{31}} + a\,O(\varepsilon^4)
\;=\; a\,O(\varepsilon)
\end{eqnarray}
Again, these shifts in the eigenvalues are negligible.
Putting everything together, we evaluate the sizes of the elements of $H''''_Y$ to find
\begin{equation}
H''''_Y = a
\left[ \begin{array}{ccc}
       O(\varepsilon^3) & O(\varepsilon^3) & 0 \\
       O(\varepsilon^3) & O(1)             & O(\varepsilon^4) \\
       0                & O(\varepsilon^4) & O(\varepsilon)
       \end{array}
\right]\;.
\end{equation}
This shows that
further diagonalization requires rotations by angles of $O(\varepsilon^3)$ or
smaller, which we will neglect.

Thus, we have found that when $\xi=O(\varepsilon^2)$ and $\delta m^2_{31}<0$
(inverted hierarchy),
we need an extra $(1,3)$-rotation to diagonalize $H$,
and the diagonalization matrix is $UVWY$.  We need to identify this product with
\begin{equation}
\tilde{U} =
\left[ \begin{array}{ccc} 1 & 0 & 0 \\
                          0 &  \tilde{c}_{23} & \tilde{s}_{23} \\
                          0 & -\tilde{s}_{23} & \tilde{c}_{23}
       \end{array}
\right]
\left[ \begin{array}{ccc} \tilde{c}_{13} & 0 & \tilde{s}_{13} e^{-i\tilde{\delta}} \\
                          0 & 1 & 0 \\
                          -\tilde{s}_{13} e^{i\tilde{\delta}} & 0 & \tilde{c}_{13}
       \end{array}
\right]
\left[ \begin{array}{ccc} \tilde{c}_{12} & \tilde{s}_{12} & 0 \\
                         -\tilde{s}_{12} & \tilde{c}_{12} & 0 \\
                          0 & 0 & 1
       \end{array}
\right]  
\end{equation}
to obtain the effective mixing angles and effective CP phase.
From Ref.~\cite{HKOT}, we know that when
$\delta m^2_{31}<0$ and $a/|\delta m^2_{31}| = O(\varepsilon^{-1})$,
identification of $\tilde{U}$ with $UVW$ leads to
\begin{eqnarray}
\tilde{\theta}_{13} & = & \theta'_{13} \;, \cr
\tilde{\theta}_{12} & = & \theta'_{12} \;, \cr
\tilde{\theta}_{23} & = & \theta_{23} \;, \cr
\tilde{\delta} & = & 0 \;.
\end{eqnarray}
Furthermore, in the range $a/|\delta m^2_{31}| = O(\varepsilon^{-1})$ we
have
\begin{eqnarray}
\theta'_{12} 
& = & \frac{\pi}{2} - \frac{\delta m^2_{21}}{2a}\sin(2\theta_{12}) + \cdots
\;=\; \frac{\pi}{2} + O(\varepsilon^3) \;,\cr
\theta'_{13} 
& = & -\frac{\delta m^2_{31}}{a}\theta_{13} + \cdots
\;=\; O(\varepsilon^2) \;,
\end{eqnarray}
which implies that
\begin{equation}
UVW \approx
\left[ \begin{array}{ccc} 1 & 0 & 0 \\
                          0 &  \tilde{c}_{23} & \tilde{s}_{23} \\
                          0 & -\tilde{s}_{23} & \tilde{c}_{23}
       \end{array}
\right]
\left[ \begin{array}{ccc} 1 & 0 & 0 \\
                          0 & 1 & 0 \\
                          0 & 0 & 1
       \end{array}
\right]
\left[ \begin{array}{ccc} 0 & 1 & 0 \\
                         -1 & 0 & 0 \\
                          0 & 0 & 1
       \end{array}
\right]  \;.
\end{equation}
Then,
\begin{eqnarray}
UVWY & \approx &
\left[ \begin{array}{ccc} 1 & 0 & 0 \\
                          0 &  \tilde{c}_{23} & \tilde{s}_{23} \\
                          0 & -\tilde{s}_{23} & \tilde{c}_{23}
       \end{array}
\right]
\left[ \begin{array}{ccc} 1 & 0 & 0 \\
                          0 & 1 & 0 \\
                          0 & 0 & 1
       \end{array}
\right]
\left[ \begin{array}{ccc} 0 & 1 & 0 \\
                         -1 & 0 & 0 \\
                          0 & 0 & 1
       \end{array}
\right]
\left[ \begin{array}{ccc} c_\eta & 0 & s_\eta \\
                          0      & 1 & 0      \\
                         -s_\eta & 0 & c_\eta
       \end{array}
\right] \cr
& \approx &
\left[ \begin{array}{ccc} 1 & 0 & 0 \\
                          0 &  \tilde{c}_{23} & \tilde{s}_{23} \\
                          0 & -\tilde{s}_{23} & \tilde{c}_{23}
       \end{array}
\right]
\left[ \begin{array}{ccc} 1 & 0 & 0 \\
						  0 & c_\eta & -s_\eta \\
                          0 & s_\eta &  c_\eta
       \end{array}
\right]\left[ \begin{array}{ccc} 1 & 0 & 0 \\
                          0 & 1 & 0 \\
                          0 & 0 & 1
       \end{array}
\right]
\left[ \begin{array}{ccc} 0 & 1 & 0 \\
                         -1 & 0 & 0 \\
                          0 & 0 & 1
       \end{array}
\right]\;.
\end{eqnarray}
Therefore, $\eta$ can be absorbed into $\tilde{\theta}_{23}$ as
\begin{eqnarray}
\tilde{\theta}_{13} & = & \theta'_{13} \;, \cr
\tilde{\theta}_{12} & = & \theta'_{12} \;, \cr
\tilde{\theta}_{23} & = & \theta_{23} - \eta\;, \cr
\tilde{\delta} & = & 0 \;.
\end{eqnarray}
As in the $\delta m^2_{31}>0$ case, the $\xi$-dependence of
$\theta'_{13} = \theta_{13} + \phi$ is very weak, so the effect of
a non-zero $\xi$ will appear dominantly in $\tilde{\theta}_{23}$.

\subsection{Summary of Neutrino Case}

To summarize what we have learned, the main effect of
including the $b\xi$ terms, which come from neutral current
universality violation, in the effective Hamiltonian is to shift
$\tilde{\theta}_{12}$ in the $\delta m^2_{31} > 0$ (normal hierarchy) case,
and $\tilde{\theta}_{23}$ in the $\delta m^2_{31} < 0$ (inverted hierarchy) case,
beyond the shifts due to the charged current interaction term $a$.
In the $\delta m^2_{31} > 0$ case, the shift in $\tilde{\theta}_{12}$ is given by
\begin{equation}
\chi \approx \dfrac{b\xi}{\delta m^2_{31}} = -\dfrac{a\xi}{2\,\delta m^2_{31}} \;,
\end{equation}
while for the $\delta m^2_{31} < 0$ case, the shift in $\tilde{\theta}_{23}$ is given by
\begin{equation}
-\eta \approx \dfrac{b\xi}{\delta m^2_{31}} 
= -\dfrac{b\xi}{|\delta m^2_{31}|} = \frac{a\xi}{2|\delta m^2_{31}|}\;.
\end{equation}
%

\section{The Effective Mixing Angles, Anti-Neutrino Case}

\subsection{Inclusion of Neutral Current Effects into the Effective Hamiltonian}

For the anti-neutrinos, the effective Hamiltonian is given by
\begin{equation}
\bar{H} = 
\edlit{U}{}^*
\left[ \begin{array}{ccc} \bar{\lambda}_1 & 0 & 0 \\
                          0 & \bar{\lambda}_2 & 0 \\
                          0 & 0 & \bar{\lambda}_3
       \end{array}
\right]
\edlit{U}{}^T
= U^*
\left[ \begin{array}{ccc} 0 & 0 & 0 \\
                          0 & \delta m^2_{21} & 0 \\
                          0 & 0 & \delta m^2_{31}
       \end{array}
\right]
U^T +
\left[ \begin{array}{ccc} -a & 0 & 0 \\
                           0 & 0 & 0 \\
                           0 & 0 & 0 
       \end{array}
\right] +
\left[ \begin{array}{ccc} -b_e &  0      &  0 \\
                           0   & -b_\mu  &  0 \\
                           0   &  0      & -b_\tau 
       \end{array}
\right] \;.
\end{equation}
The differences from the neutrino case are the reversal of signs of
the CP violating phase $\delta$ (and thus the complex conjugation of the
MNS matrix $U$), and the matter interaction terms $a$, 
and $b_\ell$ ($\ell=e,\mu,\tau$).
We denote the matter effect corrected diagonalization matrix as 
$\edlit{U}$ (note the mirror image tilde on top) to distinguish it from that
for the neutrinos.
As in the neutrino case, we make the replacement
\begin{equation}
\left[ \begin{array}{ccc} -a & 0 & 0 \\
                           0 & 0 & 0 \\
                           0 & 0 & 0 
       \end{array}
\right] +
\left[ \begin{array}{ccc} -b_e &  0     & 0 \\
                           0   & -b_\mu & 0 \\
                           0   &  0     & -b_\tau 
       \end{array}
\right] \rightarrow 
\left[ \begin{array}{ccc} -a &  0    & 0 \\
                           0 & -b\xi & 0 \\
                           0 &  0    & b\xi 
       \end{array}
\right] \;.
\end{equation}
%

\subsection{Diagonalization of the Effective Hamiltonian}

Using the matrix $Q$ from Eq.~(\ref{Qdef}),
we begin by partially diagonalize the effective Hamiltonian as
\begin{eqnarray}
\bar{H}' & = & \mathcal{Q} U^T \bar{H} U^* \mathcal{Q}^* \cr
& = &
\left[ \begin{array}{ccc} 0 & 0 & 0 \\
                          0 & \delta m^2_{21} & 0 \\
                          0 & 0 & \delta m^2_{31}
       \end{array}
\right] -
\mathcal{Q} U^T
\left[ \begin{array}{ccc} a & 0 & 0 \\
                          0 & b\xi & 0 \\
                          0 & 0 & -b\xi 
       \end{array}
\right] 
U^* \mathcal{Q}^*  \cr
& = &
\left[ \begin{array}{ccc} 0 & 0 & 0 \\
                          0 & \delta m^2_{21} & 0 \\
                          0 & 0 & \delta m^2_{31}
       \end{array}
\right] - a\,\mathcal{Q}
\left[ \begin{array}{ccc} U_{e1}U_{e1}^* & U_{e1}U_{e2}^* & U_{e1}U_{e3}^* \\
                          U_{e2}U_{e1}^* & U_{e2}U_{e2}^* & U_{e2}U_{e3}^* \\
                          U_{e3}U_{e1}^* & U_{e3}U_{e2}^* & U_{e3}U_{e3}^* 
       \end{array}
\right] \mathcal{Q}^* \cr
& &  - b\xi\,\mathcal{Q} \left\{
\left[ \begin{array}{ccc} U_{\mu 1}U_{\mu 1}^* & U_{\mu 1}U_{\mu 2}^* & U_{\mu 1}U_{\mu 3}^* \\
                          U_{\mu 2}U_{\mu 1}^* & U_{\mu 2}U_{\mu 2}^* & U_{\mu 2}U_{\mu 3}^* \\
                          U_{\mu 3}U_{\mu 1}^* & U_{\mu 3}U_{\mu 2}^* & U_{\mu 3}U_{\mu 3}^* 
       \end{array}
\right] - 
\left[ \begin{array}{ccc} U_{\tau 1}U_{\tau 1}^* & U_{\tau 1}U_{\tau 2}^* & U_{\tau 1}U_{\tau 3}^* \\
                          U_{\tau 2}U_{\tau 1}^* & U_{\tau 2}U_{\tau 2}^* & U_{\tau 2}U_{\tau 3}^* \\
                          U_{\tau 3}U_{\tau 1}^* & U_{\tau 3}U_{\tau 2}^* & U_{\tau 3}U_{\tau 3}^* 
       \end{array}
\right] \right\} \mathcal{Q}^* \;.
\end{eqnarray}
The matrix which multiplies $a$ is given by
\begin{equation}
\bar{M}_a \;=\; 
\mathcal{Q}
\left[ \begin{array}{ccc} U_{e1}U_{e1}^* & U_{e1}U_{e2}^* & U_{e1}U_{e3}^* \\
                          U_{e2}U_{e1}^* & U_{e2}U_{e2}^* & U_{e2}U_{e3}^* \\
                          U_{e3}U_{e1}^* & U{e3}U_{e2}^* & U_{e3}U_{e3}^* 
       \end{array}
\right] \mathcal{Q}^*
\;=\;
\left[ \begin{array}{ccc}
       c_{12}^2 c_{13}^2    & c_{12}s_{12}c_{13}^2 & c_{12}c_{13}s_{13} \\
       c_{12}s_{12}c_{13}^2 & s_{12}^2 c_{13}^2    & s_{12}c_{13}s_{13} \\
       c_{12}c1_{13}s_{13}   & s_{12}c_{13}s_{13}   & s_{13}^2
       \end{array}
\right]\;,
\end{equation}
while the matrix which multiplies $b\xi$ is given by
\begin{eqnarray}
\bar{M}_b & = & 
\mathcal{Q}\left\{
\left[ \begin{array}{ccc} U_{\mu 1}U_{\mu 1}^* & U_{\mu 1}U_{\mu 2}^* & U_{\mu 1}U_{\mu 3}^* \\
                          U_{\mu 2}U_{\mu 1}^* & U_{\mu 2}U_{\mu 2}^* & U_{\mu 2}U_{\mu 3}^* \\
                          U_{\mu 3}U_{\mu 1}^* & U_{\mu 3}U_{\mu 2}^* & U_{\mu 3}U_{\mu 3}^* 
       \end{array}
\right] - 
\left[ \begin{array}{ccc} U_{\tau 1}U_{\tau 1}^* & U_{\tau 1}U_{\tau 2}^* & U_{\tau 1}U_{\tau 3}^* \\
                          U_{\tau 2}U_{\tau 1}^* & U_{\tau 2}U_{\tau 2}^* & U_{\tau 2}U_{\tau 3}^* \\
                          U_{\tau 3}U_{\tau 1}^* & U_{\tau 3}U_{\tau 2}^* & U_{\tau 3}U_{\tau 3}^* 
       \end{array}
\right] \right\} \mathcal{Q}^* = M_b^* \cr
& = &
\left[ \begin{array}{l}
\phantom{-}(s_{12}^2-c_{12}^2 s_{13}^2)\cos(2\theta_{23}) 
+ \sin(2\theta_{12})\sin(2\theta_{23})s_{13}\cos\delta \\
-(1+s_{13}^2)s_{12}c_{12}\cos(2\theta_{23}) 
-(c_{12}^2 e^{+i\delta} - s_{12}^2 e^{-i\delta})s_{13}\sin(2\theta_{23}) \\
-s_{12}c_{13}\sin(2\theta_{23})e^{+i\delta} + c_{12}s_{13}c_{13}\cos(2\theta_{23}) 
\end{array} \right.  \cr
&   & \qquad\qquad
\begin{array}{l}
-(1+s_{13}^2)s_{12}c_{12}\cos(2\theta_{23}) 
-(c_{12}^2 e^{-i\delta} - s_{12}^2 e^{+i\delta})s_{13}\sin(2\theta_{23}) \\
\phantom{-}(c_{12}^2-s_{12}^2 s_{13}^2)\cos(2\theta_{23}) 
- \sin(2\theta_{12})\sin(2\theta_{23})s_{13}\cos\delta \\
\phantom{-}c_{12}c_{13}\sin(2\theta_{23})e^{+i\delta} + s_{12}s_{13}c_{13}\cos(2\theta_{23}) 
\end{array} \cr
&   & \qquad\qquad\qquad\qquad
\left. \begin{array}{l}
-s_{12}c_{13}\sin(2\theta_{23})e^{-i\delta} + c_{12}s_{13}c_{13}\cos(2\theta_{23}) \\
\phantom{-}c_{12}c_{13}\sin(2\theta_{23})e^{-i\delta} + s_{12}s_{13}c_{13}\cos(2\theta_{23}) \\
 -c_{13}^2\cos(2\theta_{23})
\end{array} \right] \;.
\end{eqnarray}
Using $\cos(2\theta_{23}) \le O(\varepsilon)$ and $\theta_{13} = O(\varepsilon)$,
the sizes of the elements of $\bar{M}_b$ are evaluated to be
\begin{equation}
\bar{M}_b = \left[ \begin{array}{ccc} O(\varepsilon) & O(\varepsilon) & O(1) \\
                              O(\varepsilon) & O(\varepsilon) & O(1) \\
                              O(1)           & O(1)           & O(\varepsilon)
             \end{array}
      \right]\;.
\end{equation}
As in the neutrino case, we neglect the $O(\varepsilon)$ terms in $\bar{M}_b$ and
approximate
\begin{equation}
\bar{M}_b \approx
\left[ \begin{array}{ccc} 0                  & 0                  & -s_{12}e^{-i\delta} \\
                          0                  & 0                  & \phantom{-}c_{12}e^{-i\delta} \\
                         -s_{12}e^{+i\delta} & c_{12}e^{+i\delta} &  0
       \end{array}
\right]\;.
\end{equation}
The effective Hamiltonian which must be diagonalized is then
\begin{eqnarray}
\bar{H}' 
& = & \mathrm{diag}(0,\delta m^2_{21},\delta m^2_{31}) - a\,\bar{M}_a - b\xi\,\bar{M}_b \cr
& = & 
\left[ 
\begin{array}{ccc}
-a\,c_{12}^2 c_{13}^2    & -a\,c_{12}s_{12}c_{13}^2 & -a\,c_{12}c_{13}s_{13}+b\xi\,s_{12}e^{-i\delta} \\
-a\,c_{12}s_{12}c_{13}^2 & -a\,s_{12}^2 c_{13}^2 + \delta m^2_{21}   & -a\,s_{12}c_{13}s_{13}-b\xi\,c_{12}e^{-i\delta} \\
-a\,c_{12}s_{13}s_{13}+b\xi\,s_{12}e^{+i\delta}   & -a\,s_{12}c_{13}s_{13}-b\xi\,c_{12}e^{+i\delta} & -a\,s_{13}^2 + \delta m^2_{31}
\end{array}
\right]\;.
\end{eqnarray}
From this point on, we set $\delta=0$ for the sake of simplicity.
Then, $\bar{H}'$ becomes
\begin{equation}
\bar{H}' =
\left[ 
\begin{array}{ccc}
-a\,c_{12}^2 c_{13}^2            & -a\,c_{12}s_{12}c_{13}^2                & -a\,c_{12}c_{13}s_{13}+b\xi\,s_{12} \\
-a\,c_{12}s_{12}c_{13}^2         & -a\,s_{12}^2 c_{13}^2 + \delta m^2_{21} & -a\,s_{12}c_{13}s_{13}-b\xi\,c_{12} \\
-a\,c_{12}s_{13}s_{13}+b\xi\,s_{12} & -a\,s_{12}c_{13}s_{13}-b\xi\,c_{12}        & -a\,s_{13}^2 + \delta m^2_{31}
\end{array}
\right] = a
\left[ \begin{array}{ccc} O(1)           & O(1)           & O(\varepsilon) \\
                          O(1)           & O(1)           & O(\varepsilon) \\
                          O(\varepsilon) & O(\varepsilon) & O(\varepsilon)
       \end{array}
\right]\
\;.
\end{equation}
%

\subsubsection{First Rotation}

Applying the Jacobi method on $\bar{H}'$, we begin by diagonalizing the 
$(1,2)$-submatrix.
Define the matrix $\bar{V}$ as
\begin{equation}
\bar{V} = 
\left[ \begin{array}{ccc} \bar{c}_{\varphi} &  \bar{s}_{\varphi} & 0 \\
	                     -\bar{s}_{\varphi} &  \bar{c}_{\varphi} & 0 \\
	                      0 & 0 & 1
	   \end{array}
\right]\;,
\label{antiVdef}
\end{equation}
where
\begin{equation}
\bar{c}_{\varphi} =\cos\bar{\varphi}\;,\quad
\bar{s}_{\varphi} =\sin\bar{\varphi}\;,\quad
\tan 2\bar{\varphi} \equiv 
-
\dfrac{a c_{13}^2\sin2\theta_{12}}{\delta m^2_{21}+a c_{13}^2\cos2\theta_{12}}\;,\quad
\left(-\theta_{12}<\bar{\varphi}\le 0\right)\;.
\label{antiphi1def}
\end{equation}
Then,
\begin{equation}
\bar{H}'' 
= \bar{V}^\dagger \bar{H}' \bar{V} 
=
\left[ \begin{array}{ccc}
       \bar{\lambda}'_1 & 0 & -a {\bar{c}}_{12}' c_{13}s_{13} + b\xi \bar{s}_{12}' \\
       0 & \bar{\lambda}'_2 & -a {\bar{s}}_{12}' c_{13}s_{13} - b\xi \bar{c}_{12}' \\
       -a {\bar{c}}_{12}' c_{13}s_{13} + b\xi \bar{s}_{12}' & -a {\bar{s}}_{12}' c_{13}s_{13} - b\xi \bar{c}_{12}' &
       -a s_{13}^2 + \delta m^2_{31}
       \end{array}
\right] \;,
\label{antiHdoubleprimedef}
\end{equation}
where
\begin{equation}
{\bar{c}}_{12}' = \cos\bar{\theta}_{12}' \;,\quad
{\bar{s}}_{12}' = \sin\bar{\theta}_{12}' \;,\quad
{\bar{\theta}}_{12}' = \theta_{12} + \bar{\varphi} \;,
\label{antitheta12primedef}
\end{equation}
and
\begin{eqnarray}
\bar{\lambda}'_1 
& = & \dfrac{(-a c_{12}^2 c_{13}^2) \bar{c}^2_\varphi
           + (a s_{12}^2 c_{13}^2 - \delta m^2_{21}) \bar{s}^2_\varphi}
            {\bar{c}^2_\varphi - \bar{s}^2_\varphi} 
\;=\; \bar{\lambda}'_{-} \;,\cr
\bar{\lambda}'_2 
& = & \dfrac{(-a s_{12}^2 c_{13}^2 + \delta m^2_{21}) \bar{c}^2_\varphi
           + (a c_{12}^2 c_{13}^2) \bar{s}^2_\varphi}
            {\bar{c}^2_\varphi - \bar{s}^2_\varphi}
\;=\; \bar{\lambda}'_{+} \;,
\label{antilambdaprimesdef}
\end{eqnarray}
with
\begin{equation}
\bar{\lambda}'_{\pm}
= \dfrac{ (\delta m^2_{21} - a c_{13}^2)
          \pm\sqrt{ (\delta m^2_{21}+a c_{13}^2)^2 - 4 a c_{13}^2 s_{12}^2 \delta m^2_{21} }
        }
        { 2 }\;.
\label{antilambdaprimeplusminusdef}
\end{equation}
From Ref.~\cite{HKOT}, we know that in the region $a/|\delta m^2_{31}| = O(\varepsilon^{-1})$,
we can expand $\bar{\varphi}$ as
\begin{equation}
\bar{\theta}_{12}' 
= \frac{\delta m_{21}^2}{2a} \sin(2\theta_{12}) + O(\varepsilon^5) 
= O(\varepsilon^3) \;.
\label{theta12primebar}
\end{equation}
Therefore,
\bea
	\bar{c}_{12}' &=& \cos\l(\frac{\delta m_{21}^2}{2a} 
	\sin(2\theta_{12})\r) = 1 - O(\varepsilon^6) \;,
\cr
	\bar{s}_{12}' &=& \sin\l(\frac{\delta m_{21}^2}{2a} 
	\sin(2\theta_{12})\r) = O(\veps^3) \;.
\eea
The expansions of $\bar{\lambda}'_\pm$ are given by
\bea
	\bar{\lambda}_-' & = & 
	-a c_{13}^2 + \delta m_{21}^2 s_{12}^2 + a\,O(\varepsilon^6)
	\;=\; a\,O(1) \;,
\cr
	\bar{\lambda}_+' & = & 
	\delta m_{21}^2 c_{12}^2 + a\,O(\varepsilon^6)
	\;=\; a\,O(\veps^3) \;.
\eea
Therefore, the sizes of the elements of $\bar{H}''$ can be evaluated to be
\begin{equation}
\bar{H}''= a
\left[ \begin{array}{ccc}
	   O(1)     & 0          & O(\veps)   \\
       0        & O(\veps^3) & O(\veps^2) \\
       O(\veps) & O(\veps^2) & O(\veps)
       \end{array}
\right] \;.
\end{equation}
As in the neutrino case, though we have a choice of whether 
we diagonalize the $(1,3)$ or the $(2,3)$ submatrix, since both require rotations by angles of $O(\varepsilon)$, we diagonalize the $(1,3)$ submatrix next to maintain the
parallel with the $\xi=0$ case.

\subsubsection{Second Rotation}

Define the matrix $\bar{W}$ as
\begin{equation}
\bar{W} = 
\left[ \begin{array}{ccc} \bar{c}_\phi & 0 & \bar{s}_\phi \\
                          0 &  1 &  0 \\
	                  -\bar{s}_\phi & 0 &  \bar{c}_\phi 
	   \end{array}
\right]\;,
\label{antiWdef}
\end{equation}
where
$\bar{c}_{\phi} = \cos\bar{\phi}$,
$\bar{s}_{\phi} = \sin\bar{\phi}$, and
\begin{eqnarray}
\tan 2\bar{\phi}
& \equiv & 
\dfrac{-2(a \bar{c}_{12}'c_{13}s_{13} - b\xi \bar{s}_{12}')}
      {\delta m^2_{31}-a s_{13}^2 - \bar{\lambda}'_1} \cr
& = & -
\dfrac{2a\bar{c}_{12}'c_{13}s_{13}}
      {\delta m^2_{31}-a s_{13}^2 - \bar{\lambda}'_1}
\left\{ 1 + \xi\left( \dfrac{ \bar{s}'_{12} }{2\bar{c}'_{12}c_{13}s_{13}} \right)
\right\} \cr
& = & \tan 2\bar{\phi}_0
\left\{ 1 + \xi\left( \dfrac{ \bar{s}'_{12} }{2\bar{c}'_{12}c_{13}s_{13}} \right)
\right\} \;.
\label{antiphi2def}
\end{eqnarray}
The angle $\bar{\phi}$ is in the fourth quadrant when 
$\delta m^2_{31} > 0$ (normal hierarchy),
and in the first quadrant when $\delta m^2_{31} < 0$ (inverted hierarchy).
$\bar{\phi}_0$ is the rotation angle when $\xi=0$.
Taking the arc-tangent of both sides of the above equation, we find
\begin{equation}
\bar{\phi} = \bar{\phi}_0
           + \xi\left(\frac{\sin 4\bar{\phi}_0 \tan\bar{\theta}'_{12}}
                           {4\sin 2\theta_{13}}
                \right)
           + O(\xi^2) \;.
\end{equation}
When $a/|\delta m^2_{31}| = O(\varepsilon^{-1})$, $\bar{\phi}_0$ is given by
\begin{equation}
\bar{\phi}_0 =
\left\{ \begin{array}{ll}
        -\theta_{13} + \dfrac{\delta m^2_{31}}{a}\,\theta_{13} + O(\varepsilon^3) &
        \qquad(\delta m^2_{31}>0) \;, \\
        \left( \dfrac{\pi}{2} - \theta_{13} \right)
        + \dfrac{\delta m^2_{31}}{a}\,\theta_{13}+ O(\varepsilon^3) &
        \qquad(\delta m^2_{31}<0) \;.
        \end{array}
\right. 
\label{phizerobar}
\end{equation}
Therefore, $\sin 4\bar{\phi}_0 \approx -\sin(4\theta_{13})$ for both the 
$\delta m^2_{31} >0$ and $\delta m^2_{31} <0$ cases and
using Eq.~(\ref{theta12primebar}), we find
\begin{equation}
\bar{\phi} \approx 
\bar{\phi}_0
- \xi \left(\dfrac{\delta m^2_{21}}{4a}\right)\sin(2\theta_{12}) 
= \bar{\phi}_0 - \frac{1}{4}O(\varepsilon^5)\;.
\end{equation}
Therefore, the difference between $\bar{\phi}$ and $\bar{\phi}_0$ can be neglected
in this range of $a$, just as in the neutrino case.

Using $\bar{W}$, we obtain
\begin{eqnarray}
\bar{H}'''
& = & \bar{W}^\dagger \bar{H}'' \bar{W} \cr
& = &
\left[ \begin{array}{ccc}
\bar{\lambda}''_1 
& (a \bar{s}_{12}' s_{13}c_{13}+b\xi \bar{c}_{12}' )\bar{s}_{\phi} 
& 0 \\
(a \bar{s}_{12}' s_{13}c_{13}+b\xi \bar{c}_{12}' )\bar{s}_{\phi} & 
\bar{\lambda}'_2 & 
-( a\bar{s}_{12}'s_{13}c_{13}+b\xi \bar{c}_{12}' )\bar{c}_\phi  \\
0 & -( a\bar{s}_{12}'s_{13}c_{13}+b\xi \bar{c}_{12}' )\bar{c}_\phi & 
\bar{\lambda}''_3
\end{array} \right]  \;,
\label{antiHtripleprimedef}
\end{eqnarray}
where
\begin{eqnarray}
\bar{\lambda}''_1 & = &  
\dfrac{ \bar{\lambda}'_1 \bar{c}^2_\phi 
+ (a s^2_{13}-\delta m^2_{31})\bar{s}^2_\phi}
{\bar{c}^2_\phi-\bar{s}^2_\phi} \;,\cr 
\bar{\lambda}''_3 & = &  
\dfrac{(-a s^2_{13}+\delta m^2_{31})\bar{c}^2_\phi 
- \bar{\lambda}'_1 \bar{s}^2_\phi}{\bar{c}^2_\phi-\bar{s}^2_\phi} \;.
\label{antilambdadoubleprimes}
\end{eqnarray}
If we define
\begin{equation}
\bar{\lambda}''_{\pm} \equiv
   \dfrac{ [(\delta m^2_{31}-a s_{13}^2)+ \bar{\lambda}'_{1}  ]
\pm \sqrt{ [(\delta m^2_{31}-a s_{13}^2)- \bar{\lambda}'_{1}  ]^2 
         + 4 ( a {\bar{c}'_{12}}s_{13}c_{13} + b\xi \bar{s}_{12}' )^2 }
         }
         { 2 } \;,
\label{antilambdadoubleprimeplusminusdef}
\end{equation}
then
\begin{eqnarray}
\bar{\lambda}''_1 = \bar{\lambda}''_{-}\;,\quad
\bar{\lambda}''_3 = \bar{\lambda}''_{+}\;,\qquad 
\mbox{if $\delta m^2_{31} > 0$}\;, \cr
\bar{\lambda}''_1 = \bar{\lambda}''_{+}\;,\quad
\bar{\lambda}''_3 = \bar{\lambda}''_{-}\;,\qquad 
\mbox{if $\delta m^2_{31} < 0$}\;.
\end{eqnarray}
%
%
%
%
%
%
When $a/|\delta m^2_{31}|=O(\varepsilon^{-1})$, we can expand 
$\bar{\lambda}_\pm''$ as
\bea
	\bar{\lambda}_-'' 
	& = & -a+s_{13}^2 \delta m_{31}^2
	+ s_{12}^2 \delta m_{21}^2
	+ O(\veps^3|\delta m_{31}^2|)
	\;=\; a\,O(1) \;,
\cr
	\bar{\lambda}_+'' 
	& = & c_{13}^2 \delta m_{31}^2
	+ O(\varepsilon^3 |\delta m_{31}^2|)
	\;=\; a\,O(\varepsilon) \;.
\eea
%
%
Also, from Eq.~(\ref{phizerobar}) and the fact that $\bar{\phi}\approx\bar{\phi}_0$, 
we conclude
\begin{equation}
\begin{array}{lll}
\bar{s}_\phi \approx -s_{13} = O(\varepsilon)\;,\quad & 
\bar{c}_\phi \approx c_{13} = O(1)\;,\quad &
(\delta m^2_{31} > 0)\;, \\
\bar{s}_\phi \approx c_{13} = O(1)\;,\quad & 
\bar{c}_\phi \approx s_{13} = O(\varepsilon)\;,\quad &
(\delta m^2_{31} < 0)\;.
\end{array}
\end{equation}
%
Putting everything together, we evaluate the sizes of the elements of 
$\bar{H}'''$ and find
\begin{equation}
\bar{H}'''= a
\left[ \begin{array}{ccc} 
       O(1)       & O(\veps^3) & 0  \\
       O(\veps^3) & O(\veps^3) & O(\veps^2) \\
       0          & O(\veps^2) & O(\veps)
\end{array}
\right] 
\end{equation}
when $\delta m_{31}^2>0$, and
\begin{equation}
\bar{H}'''= a
\left[ \begin{array}{ccc} 
	   O(\veps)   & O(\veps^2) & 0  \\
       O(\veps^2) & O(\veps^3) & O(\veps^3) \\
       0          & O(\veps^3) & O(1)
\end{array}
\right] 
\end{equation}
when $\delta m_{31}^2<0$.
Therefore, for the 
$\delta m_{31}^2>0$ (normal hierarchy) case, we must diagonalize
the $(2,3)$ submatrix next, while for the
$\delta m_{31}^2<0$ (inverted hierarchy) case, we must diagonalize the
$(1,2)$ submatrix next.

\subsubsection{Third Rotation, $\delta m^2_{31}>0$ Case}

To diagonalize the $(2,3)$ submatrix of $\bar{H}'''$, 
we define the matrix $\bar{X}$ as
\begin{equation}
\bar{X} = 
\left[ \begin{array}{ccc} 1 & 0 & 0 \\
                          0 & \bar{c}_\chi & \bar{s}_\chi \\
                          0 & -\bar{s}_\chi & \bar{c}_\chi
       \end{array}
\right] \;,
\end{equation}
where
\begin{equation}
\bar{c}_\chi = \cos\bar{\chi}\;,\quad
\bar{s}_\chi = \sin\bar{\chi}\;,\quad
\tan 2\bar{\chi} \equiv 
-\dfrac{ 2(a\bar{s}_{12}'s_{13}c_{13}+b\xi\bar{c}_{12}')\bar{c}_\phi }
       { \bar{\lambda}''_{3} - \bar{\lambda}'_{2} }\;.
\end{equation}
Then,
\begin{eqnarray}
\bar{H}''''_{X} 
& = & \bar{X}^\dagger \bar{H}''' \bar{X} \cr
& = & 
\left[ \begin{array}{ccc}
\bar{\lambda}''_1 
& (a\bar{s}_{12}'s_{13}c_{13}+b\xi\bar{c}_{12}')\bar{s}_\phi\bar{c}_\chi
& (a\bar{s}_{12}'s_{13}c_{13}+b\xi\bar{c}_{12}')\bar{s}_\phi\bar{s}_\chi \\
(a\bar{s}_{12}'s_{13}c_{13}+b\xi\bar{c}_{12}') \bar{s}_\phi\bar{c}_\chi 
& \bar{\lambda}'''_{2X} 
& 0 \\
(a\bar{s}_{12}'s_{13}c_{13}+b\xi\bar{c}_{12}') \bar{s}_\phi\bar{s}_\chi 
& 0 
& \bar{\lambda}'''_{3X}
\end{array} \right]\;, \cr
& &
\end{eqnarray}
where
\begin{eqnarray}
\bar{\lambda}'''_{2X} & = &  
\dfrac{\bar{\lambda}'_2 \bar{c}^2_\chi 
- \bar{\lambda}''_3 \bar{s}^2_\chi}{\bar{c}^2_\chi-\bar{s}^2_\chi} 
\;=\; \bar{\lambda}'''_{X-}\;,\cr 
\bar{\lambda}'''_{3X} & = &  
\dfrac{\bar{\lambda}''_3 \bar{c}^2_\chi 
- \bar{\lambda}'_2 \bar{s}^2_\chi}{\bar{c}^2_\chi-\bar{s}^2_\chi} 
\;=\; \bar{\lambda}'''_{X+}\;,
\label{antilambdatripleprimes}
\end{eqnarray}
with
\begin{equation}
\bar{\lambda}'''_{X\pm} \equiv
   \dfrac{ ( \bar{\lambda}''_{3} + \bar{\lambda}'_{2} )
\pm \sqrt{ ( \bar{\lambda}''_{3} - \bar{\lambda}'_{2} )^2 
         + 4 (a\bar{s}_{12}'s_{13}c_{13}+b\xi\bar{c}_{12}')^2 \bar{c}_\phi^2 }
         }
         { 2 } \;.
\label{antilambdatripleprimeplusminusdef}
\end{equation}
Recalling that
\begin{eqnarray}
a\bar{s}'_{12}s_{13}c_{13}
& \approx & a\left(\dfrac{\delta m^2_{21}}{2a}\sin(2\theta_{12})\right) s_{13}c_{13}
\;=\; a\,O(\varepsilon^4) \;,\cr
b\xi\bar{c}'_{12}
& \approx & -\frac{1}{2}\xi 
\;=\; -a\,O(\varepsilon^2) \;,\cr
\bar{\lambda}'_2  & \approx & \delta m^2_{21} c_{12}^2 \;=\; a\,O(\varepsilon^3)  \;,\cr
\bar{\lambda}''_3 & \approx & \delta m^2_{31} c_{13}^2 \;=\; a\,O(\varepsilon)  \;,
\end{eqnarray}
and $\bar{c}_\phi \approx 1$, we find 
\begin{equation}
\tan 2\bar{\chi} = -
\dfrac{2(a \bar{s}'_{12}s_{13}c_{13}+b\xi \bar{c}'_{12})\bar{c}_\phi}
      {\bar{\lambda}''_3 - \bar{\lambda}'_2}
= -\dfrac{2b\xi}{\delta m^2_{31}}\left\{ 1 + O(\varepsilon^2) \right\} \;.
\end{equation}
Therefore, the angle $\bar{\chi}$ is given approximately by 
\begin{equation}
\bar{\chi} \approx -\frac{b\xi}{\delta m^2_{31}} = O(\varepsilon)\;,
\end{equation}
from which we can conclude that $\bar{s}_\chi = O(\varepsilon)$ and $\bar{c}_\chi = O(1)$.
The eigenvalues can also be expanded in $\varepsilon$ and we find
\begin{eqnarray}
\bar{\lambda}'''_{2X}
& = & \delta m^2_{21} c_{12}^2 - \dfrac{b^2\xi^2}{\delta m^2_{31}} + a\,O(\varepsilon^5)
\;=\; a\,O(\varepsilon^3)\;,\cr
\bar{\lambda}'''_{3X}
& = & \delta m^2_{31} c_{13}^2 + \dfrac{b^2\xi^2}{\delta m^2_{31}} + a\,O(\varepsilon^4)
\;=\; a\,O(\varepsilon) \;.
\end{eqnarray}
As in the neutrino case, the shifts in the eigenvalues are of order $aO(\varepsilon^3)$ and
their effects on $\delta\bar{\lambda}_{31} = a O(1)$
and $\delta\bar{\lambda}_{21} = a O(1)$ are negligible.
Putting everything together, we evaluate the sizes of the elements of 
$\bar{H}''''_X$ to find
\begin{equation}
\bar{H}''''_X = a
\left[ \begin{array}{ccc}
       O(1)             & O(\varepsilon^4) & O(\varepsilon^3) \\
       O(\varepsilon^4) & O(\varepsilon^3) & 0 \\
       O(\varepsilon^3) & 0                & O(\varepsilon)
       \end{array}
\right]\;. 
\end{equation}
This shows that
further diagonalization requires rotations by angles of $O(\varepsilon^3)$ or
smaller, which we will neglect.

We have found that when $\xi=O(\varepsilon^2)$ and 
$\delta m^2_{31}>0$ (normal hierarchy), we need an extra $(2,3)$-rotation
to diagonalize $\bar{H}$, and the diagonalization matrix is
$\bar{U}\bar{V}\bar{W}\bar{X}$, which we need to identify with
\begin{equation}
\edlit{U} =
\left[ \begin{array}{ccc} 1 & 0 & 0 \\
                          0 &  \edlit{c}_{23} & \edlit{s}_{23} \\
                          0 & -\edlit{s}_{23} & \edlit{c}_{23}
       \end{array}
\right]
\left[ \begin{array}{ccc} 
\edlit{c}_{13} & 0 & \edlit{s}_{13} e^{-i\edlit{\delta}} \\
0 & 1 & 0 \\
-\edlit{s}_{13} e^{i\edlit{\delta}} & 0 & \edlit{c}_{13}
       \end{array}
\right]
\left[ \begin{array}{ccc} \edlit{c}_{12} & \edlit{s}_{12} & 0 \\
                         -\edlit{s}_{12} & \edlit{c}_{12} & 0 \\
                          0 & 0 & 1
       \end{array}
\right]  \;.
\end{equation}
From Ref.~\cite{HKOT}, we know that when $\delta m^2_{31}>0$ and
$a/|\delta m^2_{31}|=O(\varepsilon^{-1})$, identification of
$\edlit{U}$ with $\bar{U}\bar{V}\bar{W}$ yields
\begin{eqnarray}
\edlit{\theta}_{13} & \approx & 
\bar{\theta}'_{13} \;, \cr
\edlit{\theta}_{12} & \approx & 
\bar{\theta}_{12}' \;, \cr
\edlit{\theta}_{23} & \approx & \theta_{23} \;,\cr
\edlit{\delta} & \approx & 0 \;,
\end{eqnarray}
where $\bar{\theta}'_{13}\equiv \theta_{13}+\bar{\phi}$.
(Note that we are considering the $\delta=0$ case.)
Furthermore, in the range $a/|\delta m^2_{31}| = O(\varepsilon^{-1})$ we have
\begin{eqnarray}
\bar{\theta}_{12} 
& = & \dfrac{\delta m^2_{21}}{2a}\sin(2\theta_{12}) + \cdots
\;=\; O(\varepsilon^3) \;, \cr
\bar{\theta}_{13} 
& = & \dfrac{\delta m^2_{31}}{a}\theta_{13} + \cdots
\;=\; O(\varepsilon^2) \;,
\end{eqnarray}
which implies
\begin{equation}
\bar{U}\bar{V}\bar{W} \;\approx\;
\left[ \begin{array}{ccc} 1 & 0 & 0 \\
                          0 &  \edlit{c}_{23} & \edlit{s}_{23} \\
                          0 & -\edlit{s}_{23} & \edlit{c}_{23}
       \end{array}
\right]
\left[ \begin{array}{ccc} 
1 & 0 & 0 \\
0 & 1 & 0 \\
0 & 0 & 1
       \end{array}
\right]
\left[ \begin{array}{ccc} 
1 & 0 & 0 \\
0 & 1 & 0 \\
0 & 0 & 1
       \end{array}
\right]  
\;=\;
\left[ \begin{array}{ccc} 1 & 0 & 0 \\
                          0 &  \edlit{c}_{23} & \edlit{s}_{23} \\
                          0 & -\edlit{s}_{23} & \edlit{c}_{23}
       \end{array}
\right]
\;.
\end{equation}
Since $\bar{X}$ is an $(2,3)$-rotation matrix, 
multiplying $\bar{U}\bar{V}\bar{W}$ from the right with
$\bar{X}$ will only lead to a shift in $\edlit{\theta}_{23}$.
%
%
Therefore,
\begin{eqnarray}
\edlit{\theta}_{13} & \approx & \bar{\theta}'_{13} \;,\cr
\edlit{\theta}_{12} & \approx & \bar{\theta}_{12}' \;,\cr
\edlit{\theta}_{23} & \approx & \theta_{23} +\bar{\chi} \;,\cr
\edlit{\delta} & \approx & 0 \;.
\end{eqnarray}
%

\subsubsection{Third Rotation, $\delta m^2_{31}<0$ Case}

In this case, we diagonalize the $(1,2)$ submatrix of $\bar{H}'''$.
Define the matrix $\bar{Y}$ as
\begin{equation}
\bar{Y} = 
\left[ \begin{array}{ccc} \bar{c}_\eta & \bar{s}_\eta & 0 \\
                         -\bar{s}_\eta & \bar{c}_\eta & 0      \\
                          0 & 0 & 1
       \end{array}
\right] \;,
\end{equation}
where
\begin{equation}
\bar{c}_\eta = \cos\bar{\eta}\;,\quad
\bar{s}_\eta = \sin\bar{\eta}\;,\quad
\tan 2\bar{\eta} \equiv 
\dfrac{ 2(a\bar{s}_{12}'s_{13}c_{13} + b\xi\bar{c}_{12}')\bar{s}_\phi }
      { \bar{\lambda}'_{2} - \bar{\lambda}''_{1} }\;.
\end{equation}
Then,
\begin{eqnarray}
\bar{H}''''_Y 
& = & \bar{Y}^\dagger \bar{H}''' \bar{Y} \cr
& = & 
\left[ \begin{array}{ccc}
\bar{\lambda}'''_{1Y} & 0 
& (a\bar{s}_{12}'s_{13}c_{13}+b\xi\bar{c}_{12}')\bar{c}_\phi\bar{s}_\eta \\
0 & \bar{\lambda}'''_{2Y} 
& -(a\bar{s}_{12}'s_{13}c_{13}+b\xi\bar{c}_{12}')\bar{c}_\phi\bar{c}_\eta \\
(a\bar{s}_{12}'s_{13}c_{13}+b\xi\bar{c}_{12}')\bar{c}_\phi\bar{s}_\eta 
& -(a\bar{s}_{12}'s_{13}c_{13}+b\xi\bar{c}_{12}')\bar{c}_\phi\bar{c}_\eta 
& \bar{\lambda}''_3
\end{array} \right]\;, \cr
& &
\end{eqnarray}
where
\begin{eqnarray}
\bar{\lambda}'''_{1Y} & = &  
\dfrac{\bar{\lambda}''_1 \bar{c}^2_\eta 
- \bar{\lambda}'_2 \bar{s}^2_\eta}{\bar{c}^2_\eta-\bar{s}^2_\eta} 
\;=\; \bar{\lambda}'''_{Y-}\;,\cr 
\bar{\lambda}'''_{2Y} & = &  
\dfrac{\bar{\lambda}'_2 \bar{c}^2_\eta 
- \bar{\lambda}''_1 \bar{s}^2_\eta}{\bar{c}^2_\eta-\bar{s}^2_\eta} 
\;=\; \bar{\lambda}'''_{Y+}\;,
\label{antilambdaquadrupleprimes}
\end{eqnarray}
with
\begin{equation}
\bar{\lambda}'''_{Y\pm} \equiv
   \dfrac{ ( \bar{\lambda}'_{2} + \bar{\lambda}''_{1} )
\pm \sqrt{ ( \bar{\lambda}'_{2} - \bar{\lambda}''_{1} )^2 
         + 4 (a\bar{s}_{12}'s_{13}c_{13}+b\xi\bar{c}_{12}')^2 \bar{s}_\phi^2 }
         }
         { 2 } \;.
\label{antilambdaquadrupleprimeplusminusdef}
\end{equation}
Using
\begin{eqnarray}
a\bar{s}'_{12}s_{13}c_{13}
& \approx & a\left(\dfrac{\delta m^2_{21}}{2a}\sin(2\theta_{12})\right) s_{13}c_{13}
\;=\; a\,O(\varepsilon^4) \;,\cr
b\xi\bar{c}'_{12}
& \approx & -\frac{1}{2}\xi 
\;=\; -a\,O(\varepsilon^2) \;,\cr
\bar{\lambda}''_1  & \approx & c_{13}^2 \delta m^2_{31} \;=\; a\,O(\varepsilon)   \;,\cr
\bar{\lambda}'_2   & \approx & c_{12}^2 \delta m^2_{21} \;=\; a\,O(\varepsilon^3) \;,
\end{eqnarray}
and $\bar{s}_\phi \approx 1$, we find
\begin{equation}
\tan 2\bar{\eta} = 
\dfrac{2(a \bar{s}'_{12}s_{13}c_{13}+b\xi \bar{c}'_{12})\bar{s}_\phi}
      {\bar{\lambda}'_2 - \bar{\lambda}''_1}
= -\dfrac{2b\xi}{\delta m^2_{31}}\left\{ 1 + O(\varepsilon^2) \right\} \;.
\end{equation}
Therefore, the angle $\bar{\eta}$ is given approximately by 
\begin{equation}
\bar{\eta} \approx -\frac{b\xi}{\delta m^2_{31}}
= \frac{b\xi}{|\delta m^2_{31}|} = O(\varepsilon)\;,
\end{equation}
from which we can conclude that $\bar{s}_\eta = O(\varepsilon)$ and $\bar{c}_\eta = O(1)$.
The eigenvalues can also be expanded in $\varepsilon$ and we find
\begin{eqnarray}
\bar{\lambda}'''_{1Y}
& = & \delta m^2_{31} c_{13}^2 + \dfrac{b^2\xi^2}{\delta m^2_{31}} + a\,O(\varepsilon^4)
\;=\; a\,O(\varepsilon) \;,\cr
\bar{\lambda}'''_{2Y}
& = & \delta m^2_{21} c_{12}^2 - \dfrac{b^2\xi^2}{\delta m^2_{31}} + a\,O(\varepsilon^5)
\;=\; a\,O(\varepsilon^3) \;.
\end{eqnarray}
Again, the shifts are negligible.
Putting everything together, we evaluate the sizes of the elements of 
$\bar{H}''''_Y$ to find
\begin{equation}
\bar{H}''''_Y = a
\left[ \begin{array}{ccc}
       O(\varepsilon)   & 0                & O(\varepsilon^4) \\
       0                & O(\varepsilon^3) & O(\varepsilon^3) \\
       O(\varepsilon^4) & O(\varepsilon^3) & O(1)
       \end{array}
\right]\;,
\end{equation}
which shows that
further diagonalization requires rotations by angles of $O(\varepsilon^3)$ or
smaller, which we will neglect.

Thus, we have found that when $\xi=O(\varepsilon^2)$ and $\delta m^2_{31}<0$,
the diagonalization of $\bar{H}$ requires an
extra $(1,2)$ rotation, and the diagonalization matrix is
$\bar{U}\bar{V}\bar{W}\bar{Y}$.
As in the $\delta m^2_{31}>0$ case, $\bar{U}\bar{V}\bar{W}\bar{Y}$ must be
identified with
\begin{equation}
\edlit{U} =
\left[ \begin{array}{ccc} 1 & 0 & 0 \\
                          0 &  \edlit{c}_{23} & \edlit{s}_{23} \\
                          0 & -\edlit{s}_{23} & \edlit{c}_{23}
       \end{array}
\right]
\left[ \begin{array}{ccc} \edlit{c}_{13} & 0 & \edlit{s}_{13} e^{-i\edlit{\delta}} \\
                          0 & 1 & 0 \\
                          -\edlit{s}_{13} e^{i\edlit{\delta}} & 0 & \edlit{c}_{13}
       \end{array}
\right]
\left[ \begin{array}{ccc} \edlit{c}_{12} & \edlit{s}_{12} & 0 \\
                         -\edlit{s}_{12} & \edlit{c}_{12} & 0 \\
                          0 & 0 & 1
       \end{array}
\right]  \;.
\end{equation}
Again, from Ref.~\cite{HKOT}, we know that
the identification of $\bar{U}\bar{V}\bar{W}$ with $\edlit{U}$ yields
\begin{eqnarray}
\edlit{\theta}_{13} & \approx & \bar{\theta}'_{13} \;, \cr
\edlit{\theta}_{12} & \approx & \frac{c_{13}}{\bar{c}_{13}'}
\l( \frac{\delta m_{21}^2}{2a} \r) \sin(2\theta_{12}) \;, \cr
\edlit{\theta}_{23} & \approx & \theta_{23}
-\frac{\bar{s}_\phi}{\bar{c}_{13}'}
\l( \frac{\delta m_{21}^2}{2a} \r)\sin(2\theta_{12}) \;, \cr
\edlit{\delta} & \approx & 0 \;,
\end{eqnarray}
where
$\bar{c}'_{13} = \cos\bar{\theta}'_{13}$.
Since $\bar{Y}$ is an $(1,2)$-rotation matrix,
multiplying $\bar{U}\bar{V}\bar{W}$ from the right by $\bar{Y}$ will only 
lead to a shift in $\edlit{\theta}_{12}$.
Therefore,
\begin{eqnarray}
\edlit{\theta}_{13} & \approx & \bar{\theta}'_{13} \;, \cr
\edlit{\theta}_{12} & \approx & \frac{c_{13}}{\bar{c}_{13}'}
\l( \frac{\delta m_{21}^2}{2a} \r) \sin(2\theta_{12}) + \bar{\eta} \;, \cr
\edlit{\theta}_{23} & \approx & \theta_{23}
-\frac{\bar{s}_\phi}{\bar{c}_{13}'}
\l( \frac{\delta m_{21}^2}{2a} \r)\sin(2\theta_{12}) \;, \cr
\edlit{\delta} & \approx & 0 \;.
\end{eqnarray}
%

\subsection{Summary of Anti-Neutrino Case}

To summarize, in contrast to the neutrino case,
the main effect of including the $b\xi$ terms in the effective Hamiltonian 
for the anti-neutrinos is to shift
$\edlit{\theta}_{23}$ in the $\delta m^2_{31} > 0$ (normal hierarchy) case,
and $\edlit{\theta}_{12}$ in the $\delta m^2_{31} < 0$ (inverted hierarchy) case.
The mixing angle that is affected depending on the sign of $\delta m^2_{31}$ is
the exact opposite of the neutrino case.
In the $\delta m^2_{31} > 0$ case, the shift in $\edlit{\theta}_{23}$ is given by
\begin{equation}
\bar{\chi} \approx -\dfrac{b\xi}{\delta m^2_{31}} = \dfrac{a\xi}{2\,\delta m^2_{31}} \;,
\end{equation}
while for the $\delta m^2_{31} < 0$ case, the shift in $\edlit{\theta}_{12}$ is given by
\begin{equation}
\bar{\eta} \approx -\dfrac{b\xi}{\delta m^2_{31}} 
= \dfrac{b\xi}{|\delta m^2_{31}|} = -\dfrac{a\xi}{2|\delta m^2_{31}|} \;.
\end{equation}

Listing these results together with those for the neutrino case from
the previous section, we obtain Table~\ref{EffectOfNCUV}.  
The accuracy of our approximation will be demonstrated
later by comparing our conclusions with the exact numerical results.
Let us now investigate how these shifts in the effective mixing angles 
affect the oscillation probabilities.


\begin{table}
\begin{tabular}{|c||c|c|}
\hline
& \ $\delta m^2_{31}>0$ (normal hierarchy)   \ 
& \ $\delta m^2_{31}<0$ (inverted hierarchy) \ \\
\hline\hline
\ Neutrino\ \      
& \ $\tilde{\theta}_{12}$ is shifted by $-\dfrac{a\xi}{2\,\delta m^2_{31}}$   \ 
& \ $\tilde{\theta}_{23}$ is shifted by $+\dfrac{a\xi}{2| \delta m^2_{31}|}$ \ \\
\hline
\ Anti-neutrino\ \ 
& \ $\edlit{\theta}_{23}$ is shifted by $+\dfrac{a\xi}{2\,\delta m^2_{31}}$   \
& \ $\edlit{\theta}_{12}$ is shifted by $-\dfrac{a\xi}{2 |\delta m^2_{31}|}$ \ \\
\hline
\end{tabular}
\caption{Matter effects from neutral current universality violation.
The parameter $\xi$, defined in Eq.~(\protect{\ref{xidef}}), gives the size of the violation.}
\label{EffectOfNCUV}
\end{table}

\section{The Oscillation Probabilities}

The oscillation probability from neutrino flavor $\nu_\alpha$ to
neutrino flavor $\nu_\beta$ in vacuum is given by
\begin{eqnarray}
P(\nu_\alpha\rightarrow\nu_\alpha)
& = & 1 - 4\, |U_{\alpha 2}|^2 \left( 1 - |U_{\alpha 2}|^2 \right)
            \sin^2\frac{\Delta_{21}}{2}
        - 4\, |U_{\alpha 3}|^2 \left( 1 - |U_{\alpha 3}|^2 \right)
            \sin^2\frac{\Delta_{31}}{2} \cr
& &  \phantom{1}
        + 2\, |U_{\alpha 2}|^2 |U_{\alpha 3}|^2
          \left( 4\sin^2\frac{\Delta_{21}}{2}\sin^2\frac{\Delta_{31}}{2}
                + \sin\Delta_{21}\sin\Delta_{31}
          \right) \;,
\label{Palphatoalpha}
\end{eqnarray}
for the $\alpha=\beta$ case, and
\begin{eqnarray}
P(\nu_\alpha \rightarrow \nu_\beta)
& = & 4\, |U_{\alpha 2}|^2 |U_{\beta 2}|^2 \sin^2\frac{\Delta_{21}}{2}
     +4\, |U_{\alpha 3}|^2 |U_{\beta 3}|^2 \sin^2\frac{\Delta_{31}}{2} \cr
&   & +2\,\Re( U^*_{\alpha 3}U_{\beta 3}U_{\alpha 2}U^*_{\beta 2})
      \left(4\sin^2\frac{\Delta_{21}}{2}\sin^2\frac{\Delta_{31}}{2}
           +\sin\Delta_{21}\sin\Delta_{31}
      \right) \cr 
&   & +4\,J_{(\alpha,\beta)}
      \left( \sin^2\frac{\Delta_{21}}{2}\sin\Delta_{31}
            -\sin^2\frac{\Delta_{31}}{2}\sin\Delta_{21}
      \right) \;,    
\label{Palphatonotalpha}       
\end{eqnarray}
for the $\alpha\neq\beta$ case, 
where $J_{(\alpha,\beta)}$ is the Jarskog invariant,
\begin{eqnarray}
J_{(\alpha,\beta)}
& = & +\Im(U^*_{\alpha 1}U_{\beta 1}U_{\alpha 2}U^*_{\beta 2})
\;=\; +\Im(U^*_{\alpha 2}U_{\beta 2}U_{\alpha 3}U^*_{\beta 3})
\;=\; +\Im(U^*_{\alpha 3}U_{\beta 3}U_{\alpha 1}U^*_{\beta 1}) \cr
& = & -\Im(U^*_{\alpha 2}U_{\beta 2}U_{\alpha 1}U^*_{\beta 1})
\;=\; -\Im(U^*_{\alpha 1}U_{\beta 1}U_{\alpha 3}U^*_{\beta 3})
\;=\; -\Im(U^*_{\alpha 3}U_{\beta 3}U_{\alpha 2}U^*_{\beta 2}) \cr
& = & -J_{(\beta,\alpha)}\;,
\end{eqnarray}
and 
\begin{equation}
\Delta_{ij} 
\;\equiv\; \dfrac{\delta m_{ij}^2}{2E} L
\;=\; 2.534\;\dfrac{(\delta m_{ij}^2\,/\mathrm{eV}^2)}{(E\,/\mathrm{GeV})}\,(L\,/\mathrm{km})
\;,\qquad
\delta m_{ij}^2
\equiv m_i^2-m_j^2\;.
\end{equation}
The oscillation probabilities for anti-neutrinos can be obtained by
replacing $U$ with its complex conjugate, which amounts to flipping the
sign of the CP violating phase $\delta$.

The oscillation probabilities in matter are obtained by
making the replacements
\begin{equation}
U_{\alpha i}\;\rightarrow\;\tilde{U}_{\alpha i}\;,\qquad
\Delta_{ij}\;\rightarrow\;\tilde{\Delta}_{ij} = \dfrac{\lambda_i - \lambda_j}{2E} L \;,
\end{equation}
for the neutrinos, and
\begin{equation}
U_{\alpha i}\;\rightarrow\;\edlit{U}_{\alpha i}\;,\qquad
\Delta_{ij}\;\rightarrow\;\edlit{\Delta}_{ij} = \dfrac{\bar{\lambda}_i - \bar{\lambda}_j}{2E} L \;,
\end{equation}
for the anti-neutrinos.  For instance, the $\nu_\mu$ and $\bar{\nu}_\mu$
survival probabilities in matter are given by
\begin{eqnarray}
\tilde{P}({\nu}_\mu\rightarrow{\nu}_\mu) & = &
1 
- 4\,|\tilde{U}_{\mu 2}|^2 \left( 1- |\tilde{U}_{\mu 2}|^2 \right) 
\sin^2\frac{\tilde{\Delta}_{21}}{2}
- 4\,|\tilde{U}_{\mu 3}|^2 \left( 1- |\tilde{U}_{\mu 3}|^2 \right) 
\sin^2\frac{\tilde{\Delta}_{31}}{2} \cr
& & \phantom{1}
+ 2\,|\tilde{U}_{\mu 2}|^2 |\tilde{U}_{\mu 3}|^2
\left( 4\sin^2\dfrac{\tilde{\Delta}_{21}}{2}\sin^2\dfrac{\tilde{\Delta}_{31}}{2}
      + \sin\tilde{\Delta}_{21} \sin\tilde{\Delta}_{31} \right) \;, 
\label{Pnumu2numu} \\
\edlit{P}(\bar{\nu}_\mu\rightarrow\bar{\nu}_\mu) & = &
1 
- 4\,|\edlit{U}_{\mu 2}|^2 \left( 1- |\edlit{U}_{\mu 2}|^2 \right) 
\sin^2\frac{\edlit{\Delta}_{21}}{2}
- 4\,|\edlit{U}_{\mu 3}|^2 \left( 1- |\edlit{U}_{\mu 3}|^2 \right) 
\sin^2\frac{\edlit{\Delta}_{31}}{2} \cr
& & \phantom{1}
+ 2\,|\edlit{U}_{\mu 2}|^2 |\edlit{U}_{\mu 3}|^2
\left( 4\sin^2\dfrac{\edlit{\Delta}_{21}}{2}\sin^2\dfrac{\edlit{\Delta}_{31}}{2}
      + \sin\edlit{\Delta}_{21} \sin\edlit{\Delta}_{31} \right) \;.
\label{Pnumubar2numubar}
\end{eqnarray}
Let us calculate these probabilities in the range
$a/|\delta m^2_{31}|=O(\varepsilon^{-1})$, using the results of the
previous sections.

\subsection{Neutrino Oscillations}

\subsubsection{$\delta m^2_{31}>0$ Case}

From Eq.~(\ref{Pnumu2numu}), we note that we need $\tilde{U}_{\mu 2}$
and $\tilde{U}_{\mu 3}$ to calculate $\tilde{P}(\nu_\mu\rightarrow \nu_\mu)$.
When $\delta = 0$, these are given by
\begin{eqnarray}
\tilde{U}_{\mu 2} & = & \tilde{c}_{12}\tilde{c}_{23}-\tilde{s}_{12}\tilde{s}_{13}\tilde{s}_{23} \;,\cr
\tilde{U}_{\mu 3} & = & \tilde{c}_{13}\tilde{s}_{23} \;.
\end{eqnarray}
For the $\delta m^2_{31}>0$ case,
the effective mixing angles in the region $a/|\delta m^2_{31}|=O(\varepsilon^{-1})$
are well approximated by \cite{HKOT}
\begin{eqnarray}
\tilde{\theta}_{13} 
& \approx & \frac{\pi}{2} - \left(\frac{\delta m^2_{31}}{2a}\right)\sin(2\theta_{13})
\;=\; \frac{\pi}{2} - O(\varepsilon^2)
\;, \cr
\tilde{\theta}_{12} & \approx & \frac{\pi}{2} - \frac{c_{13}}{c'_{13}}
\left( \frac{\delta m^2_{21}}{2a}\right) \sin(2\theta_{12}) 
+ \chi
\;=\; \frac{\pi}{2} - O(\varepsilon) 
\;, \cr
\tilde{\theta}_{23} & \approx & \theta_{23} + \frac{s_\phi}{c'_{13}}
\left( \frac{\delta m^2_{21}}{2a}\right) \sin(2\theta_{12}) 
\;=\; \theta_{23} + O(\varepsilon)
\;.
\end{eqnarray}
Using $\tilde{s}_{13} = 1-O(\varepsilon^4)$, 
$\tilde{c}_{13} = O(\varepsilon^2)$, we find
\begin{eqnarray}
\tilde{U}_{\mu 2} & \approx & 
\tilde{c}_{12}\tilde{c}_{23}-\tilde{s}_{12}\tilde{s}_{23}
\;=\; \cos(\tilde{\theta}_{12} + \tilde{\theta}_{23}) \;, \cr
\tilde{U}_{\mu 3} & \approx & 0  \;.
\end{eqnarray}
Therefore,
\[
\tilde{P}({\nu}_\mu\rightarrow{\nu}_\mu)
\approx 1 
- \sin^2\{2(\tilde{\theta}_{12} + \tilde{\theta}_{23})\} 
\sin^2\frac{\tilde{\Delta}_{21}}{2} \;.
\]
Note that
\begin{equation}
\tilde{\theta}_{12} + \tilde{\theta}_{23}
\approx \frac{\pi}{2} + \theta_{23}
+ \left(\frac{s_\phi - c_{13}}{c'_{13}}\right)\left(\frac{\delta m^2_{21}}{2a}\right)
  \sin(2\theta_{12}) 
+ \chi\;.
\end{equation}
Using $s'_{13}= 1-O(\varepsilon^4)$, 
$c'_{13}= O(\varepsilon^2)$, we find
\begin{equation}
s_\phi = \sin(\theta'_{13}-\theta_{13})
= s'_{13}c_{13}+c'_{13}s_{13}
\approx c_{13} \;,
\end{equation}
which shows that the $O(\varepsilon)$ terms in $\tilde{\theta}_{12}$ and
$\tilde{\theta}_{23}$ other than $\chi$ cancel (this only happens for the $\delta=0$ case
considered here) and we can approximate
\begin{equation}
\tilde{\theta}_{12} + \tilde{\theta}_{23} \approx \frac{\pi}{2} + \theta_{23}+ \chi\;.
\end{equation}
Therefore,
\begin{equation}
\tilde{P}({\nu}_\mu\rightarrow{\nu}_\mu)
\approx 1 
- \sin^2\{2(\theta_{23}+\chi)\} 
\sin^2\frac{\tilde{\Delta}_{21}}{2} \;.
\end{equation}
%

%

\subsubsection{$\delta m^2_{31}<0$ Case}

For the $\delta m^2_{31}<0$ case, the effective mixing angles
in the region $a/|\delta m^2_{31}|=O(\varepsilon^{-1})$ 
are well approximated by \cite{HKOT}
\begin{eqnarray}
\tilde{\theta}_{13} 
& \approx & -\left(\frac{\delta m^2_{31}}{a}\right)\theta_{13}
\;=\; O(\varepsilon^2)
\;, \cr
\tilde{\theta}_{12} & \approx & \frac{\pi}{2} - 
\left( \frac{\delta m^2_{21}}{2a}\right) \sin(2\theta_{12}) 
\;=\; \frac{\pi}{2} - O(\varepsilon^3)
\;, \cr
\tilde{\theta}_{23} & \approx & \theta_{23} - \eta \;.
\end{eqnarray}
Using $\tilde{s}_{13} = O(\varepsilon^2)$, 
$\tilde{c}_{13} = 1-O(\varepsilon^4)$,
$\tilde{s}_{12} = 1-O(\varepsilon^6)$, 
$\tilde{c}_{12} = O(\varepsilon^3)$, we find
\begin{eqnarray}
\tilde{U}_{\mu 2} & \approx & 0
\;, \cr
\tilde{U}_{\mu 3} & \approx & \tilde{s}_{23}  \;.
\end{eqnarray}
Therefore,
\begin{eqnarray}
\tilde{P}({\nu}_\mu\rightarrow{\nu}_\mu)
& \approx & 1 - \sin^2 (2\tilde{\theta}_{23}) 
\sin^2\frac{\tilde{\Delta}_{31}}{2} \cr
& \approx & 1 - \sin^2\{2(\theta_{23}-\eta)\}
\sin^2\frac{\tilde{\Delta}_{31}}{2}
\;.
\end{eqnarray}
%

\subsection{Anti-Neutrino Oscillations}

\subsubsection{$\delta m^2_{31}>0$ Case}
To calculate the $\bar{\nu}_\mu$ survival probability, we need
$\edlit{U}_{\mu 2}$ and $\edlit{U}_{\mu 3}$ 
as can be seen from Eq.~(\ref{Pnumubar2numubar}).
When $\delta=0$, we have
\begin{eqnarray}
\edlit{U}_{\mu 2} & = & \edlit{c}_{12}\edlit{c}_{23}-\edlit{s}_{12}\edlit{s}_{13}\edlit{s}_{23} \;,\cr
\edlit{U}_{\mu 3} & = & \edlit{c}_{13}\edlit{s}_{23} \;.
\end{eqnarray}
For the $\delta m^2_{31}>0$ case, the effective mixing angles  
in the region $a/|\delta m^2_{31}|=O(\varepsilon^{-1})$
are well approximated by \cite{HKOT}
\begin{eqnarray}
\edlit{\theta}_{12} & \approx & \dfrac{\delta m^2_{21}}{2a}\sin(2\theta_{12})
\;=\; O(\varepsilon^3) \;, \cr
\edlit{\theta}_{13} & \approx & \dfrac{\delta m^2_{31}}{a}\theta_{13}
\;=\; O(\varepsilon^2) \;, \cr
\edlit{\theta}_{23} & \approx & \theta_{23} + \bar{\chi} \;.
\end{eqnarray}
Therefore, $\edlit{s}_{12} = O(\varepsilon^3)$, $\edlit{c}_{12} = 1-O(\varepsilon^6)$,
$\edlit{s}_{13} = O(\varepsilon^2)$, $\edlit{c}_{13} = 1-O(\varepsilon^4)$, and
we can approximate
\begin{eqnarray}
\edlit{U}_{\mu 2} & \approx & \edlit{c}_{23} \;, \cr
\edlit{U}_{\mu 3} & \approx & \edlit{s}_{23} \;,
\end{eqnarray}
which yields
\begin{eqnarray}
\edlit{P}(\bar{\nu}_\mu\rightarrow\bar{\nu}_\mu)
& \approx &
1 - \sin^2(2\edlit{\theta}_{23}) \sin^2\frac{\edlit{\Delta}_{32}}{2} \cr
& \approx &
1 - \sin^2\{2(\theta_{23}+\bar{\chi})\} \sin^2\frac{\edlit{\Delta}_{32}}{2} \;.
\end{eqnarray}
%

\subsubsection{$\delta m^2_{31}<0$ Case}

For the $\delta m^2_{31}<0$ case,
the effective mixing angles in the region $a/|\delta m^2_{31}|=O(\varepsilon^{-1})$
are well approximated by \cite{HKOT}
\begin{eqnarray}
\edlit{\theta}_{13} & \approx & 
\frac{\pi}{2} + \left(\frac{\delta m^2_{31}}{a}\right)\theta_{13}
\;=\;\frac{\pi}{2}+O(\varepsilon^2) \;, \cr
\edlit{\theta}_{12} & \approx & 
\frac{c_{13}}{\bar{c}'_{13}}
\left( \frac{\delta m^2_{21}}{2a}\right) \sin(2\theta_{12}) + \bar{\eta}
\;=\; O(\varepsilon) \;, \cr
\edlit{\theta}_{23} & \approx & 
\theta_{23} - \frac{\bar{s}_\phi}{\bar{c}'_{13}}
\left( \frac{\delta m^2_{21}}{2a}\right) \sin(2\theta_{12}) 
\;=\;\theta_{23} + O(\varepsilon)
\;. 
\end{eqnarray}
Using 
$\edlit{s}_{13} = 1-O(\varepsilon^4)$,
$\edlit{c}_{13} = O(\varepsilon^2)$,
we can approximate
\begin{eqnarray}
\edlit{U}_{\mu 2} & \approx & 
\edlit{c}_{12}\edlit{c}_{23}-\edlit{s}_{12}\edlit{s}_{23}
=\cos(\edlit{\theta}_{12}+\edlit{\theta}_{23})
\;, \cr
\edlit{U}_{\mu 3} & \approx & 0 \;,
\end{eqnarray}
which yields
\begin{equation}
\edlit{P}(\bar{\nu}_\mu\rightarrow\bar{\nu}_\mu)
\approx 1 
- \sin^2\{2(\edlit{\theta}_{12}+\edlit{\theta}_{23})\} 
\sin^2\frac{\edlit{\Delta}_{21}}{2} \;.
\end{equation}
Note that
\begin{equation}
\edlit{\theta}_{12} + \edlit{\theta}_{23}
\approx \theta_{23} -
\left(\frac{\bar{s}_\phi - c_{13} }{\bar{c}'_{13}}\right)
\left(\frac{\delta m^2_{21}}{2a}\right) \sin(2\theta_{12}) + \bar{\eta} \;.
\end{equation}
Using $\bar{s}'_{13} = 1-O(\varepsilon^4)$, $\bar{c}'_{13}=O(\varepsilon^2)$, we find
\begin{equation}
\bar{s}_\phi 
\;=\; \sin(\bar{\theta}'_{13}-\theta_{13})
\;=\; \bar{s}'_{13}c_{13} - \bar{c}'_{13}s_{13}
\;\approx\; c_{13}\;,
\end{equation}
which allows us to approximate
\begin{equation}
\edlit{\theta}_{12} + \edlit{\theta}_{23}  
\;\approx\; \theta_{23} + \bar{\eta}\;.
\end{equation}
(Again, the cancellation of the $O(\varepsilon)$ terms other than $\bar{\eta}$ occurs
only for the $\delta=0$ case considered here.)
Therefore,
\begin{equation}
\edlit{P}(\bar{\nu}_\mu\rightarrow\bar{\nu}_\mu)
\approx 1 
- \sin^2\{2(\theta_{23} + \bar{\eta})\} 
\sin^2\frac{\edlit{\Delta}_{21}}{2} \;.
\end{equation}
%

\subsection{Summary of Oscillation Probabilities}

To summarize what we have found, the $\nu_\mu$ and $\bar{\nu}_\mu$
survival probabilities for the $\delta m^2_{31}>0$ (normal hierarchy) case
are given by
\begin{eqnarray}
\tilde{P}({\nu}_\mu\rightarrow{\nu}_\mu)
& \approx & 1 - \sin^2\{2(\theta_{23} - \zeta)\} 
\sin^2\frac{\tilde{\Delta}_{21}}{2} \;, \cr
\edlit{P}(\bar{\nu}_\mu\rightarrow\bar{\nu}_\mu)
& \approx & 1 - \sin^2\{2(\theta_{23} + \zeta)\} 
\sin^2\frac{\edlit{\Delta}_{32}}{2} \;,
\label{Pnumu2numuNormal}
\end{eqnarray}
while for the $\delta m^2_{31}<0$ (inverted hierarchy) case, they are given by
\begin{eqnarray}
\tilde{P}({\nu}_\mu\rightarrow{\nu}_\mu)
& \approx & 1 - \sin^2\{2(\theta_{23} + \zeta)\}
\sin^2\frac{\tilde{\Delta}_{31}}{2} \;, \cr
\edlit{P}(\bar{\nu}_\mu\rightarrow\bar{\nu}_\mu)
& \approx & 1 - \sin^2\{2(\theta_{23} - \zeta)\} 
\sin^2\frac{\edlit{\Delta}_{21}}{2} \;,
\label{Pnumu2numuInverted}
\end{eqnarray}
where we have defined
\begin{equation}
\zeta \equiv \dfrac{a\xi}{2|\delta m^2_{31}|} \;.
\end{equation}
Therefore,
though the effect of a non-zero $\xi$ appears in different effective mixing angles
depending on the mass hierarchy, and whether the particle considered
is the neutrino or the anti-neutrino (cf. Table~\ref{EffectOfNCUV}),
the net effect on the $\nu_\mu$ and $\bar{\nu}_\mu$ survival
probabilities for all cases is to shift $\theta_{23}$ in the oscillation amplitude.

Unfortunately, this shift in $\theta_{23}$ may be difficult to observe. 
The current experimentally preferred value of 
$\sin^2(2\theta_\mathrm{atm})\approx\sin^2(2\theta_{23})$ is one, with
the 90\% lower limit given by \cite{atmos,K2K}
\begin{equation}
\sin^2(2\theta_{23}) > 0.92\;.
\end{equation}
Given the shape of the function $\sin^2(2\theta_{23})$ around $\theta_{23}=\pi/4$,
$\sin^2(2\theta_{23})$ is insensitive to small shifts in $\theta_{23}$.
Indeed, because of this, the angle $\theta_{23}$ itself is ill constrained, the above limit translating into
\begin{equation}
\theta_{23} = (0.2\sim 0.3)\pi\;.
\end{equation}

However, our knowledge of the value of $\sin^2(2\theta_{23})$ is to be
improved considerably in the near future.
The long baseline neutrino oscillation experiments MINOS \cite{MINOS}, T2K \cite{T2K}, NO$\nu$A \cite{NOvA}, and others \cite{Diwan,T2B,H2B,Antusch:2004yx}
will measure $\sin^2(2\theta_\mathrm{atm}) = 4|U_{\mu 3}|^2 (1-|U_{\mu 3}|^2) = 4s_{23}^2 c_{13}^2 (1-s_{23}^2 c_{13}^2)$ from $\nu_\mu\rightarrow \nu_\mu$
to better than 1\%, 
while the reactor neutrino experiments Double-Chooz \cite{DoubleChooz}, KASKA \cite{kaska}, 
Braidwood \cite{Braidwood}, etc. \cite{Anderson:2004pk}
are expected to measure $\sin^2(2\theta_\mathrm{rct}) = 4|U_{e3}|^2 (1-|U_{e3}|^2) = \sin^2(2\theta_{13})$ from $\bar{\nu}_e\rightarrow\bar{\nu}_e$ to an accuracy of $\pm 0.01$.
These developments combined will determine $\sin^2(2\theta_{23})$ to better than 1\%,
albeit with a two-fold degeneracy. 
This degeneracy can be broken, in principle, by determining 
$4 |U_{\mu 3}|^2 |U_{e 3}|^2 = s_{23}^2 \sin^2(2\theta_{13})$ from the 
CP non-violating part of the $\nu_\mu\rightarrow\nu_e$ oscillation probability 
\cite{NOvA,SuperNOvA,Minakata:1997td,Barger:2002xk,Minakata:2002jv,Aoki:2003kc,McConnel:2004bd,Huber:2004ug,Ishitsuka:2005qi,Hagiwara:2005pe,Hiraide:2006vh}.
Therefore, a unique and accurate value of $\sin^2(2\theta_{23})$, together with
whether $\theta_{23}$ is larger or smaller than $\pi/4$, may be known. 
Furthermore, if the 1-Megaton Hyper-Kamiokande (HyperK) detector is ever constructed,
a JPARC$\rightarrow$HyperK long-baseline experiment will improve the limits even further 
\cite{HyperK}.

Even then, if the central value of $\sin^2(2\theta_{23})$ is too close to one, then the
shift due to $\zeta$ will be invisible.
Let us assume, for the sake of argument, that a 1\% shift in $\sin^2(2\theta_{23})$ is detectable.
Since
\begin{equation}
\sin^2\{2(\theta_{23}\pm\zeta)\}
= \sin^2(2\theta_{23}) \pm 2\sin(4\theta_{23})\zeta\;,
\end{equation}
the shift due to $\zeta$ would be visible if
\begin{equation}
|2\sin(4\theta_{23})\zeta| > 0.01\;.
\end{equation}
The size of $\zeta$ for $\rho=4.6\,\mathrm{g/cm^3}$, 
$E=17\,\mathrm{GeV}$, $\xi=0.025$, and $|\delta m^2_{31}|=2.5\times 10^{-3}\,\mathrm{eV}^2$,
for instance, is
\begin{equation}
\zeta = 0.03 \;.
\end{equation}
For this shift to be visible, we must have
\begin{equation}
|\sin(4\theta_{23})| > \frac{1}{6}\;,
\end{equation}
or
\begin{equation}
\sin^2(2\theta_{23}) < 0.993\;.
\end{equation}
If we require a 2\% shift, the limit will be $\sin^2(2\theta_{23}) < 0.97$, and a
3\% shift would require $\sin^2(2\theta_{23}) < 0.93$.
Therefore, whether the effect we are considering can be observed or not
depends crucially on the value of $\sin^2(2\theta_{23})$.

\section{Numerical Results}

The discussions up to this point were all based on approximate analytical calculations. To illustrate the accuracy of our analytical results, we
presenting here the results of a numerical calculation of the
effective mass-squared-differences, effective mixing angles, and oscillation 
probabilities.

\begin{figure}[p]
\begin{center}
\[
\begin{array}{rr}
\includegraphics[scale=0.36]{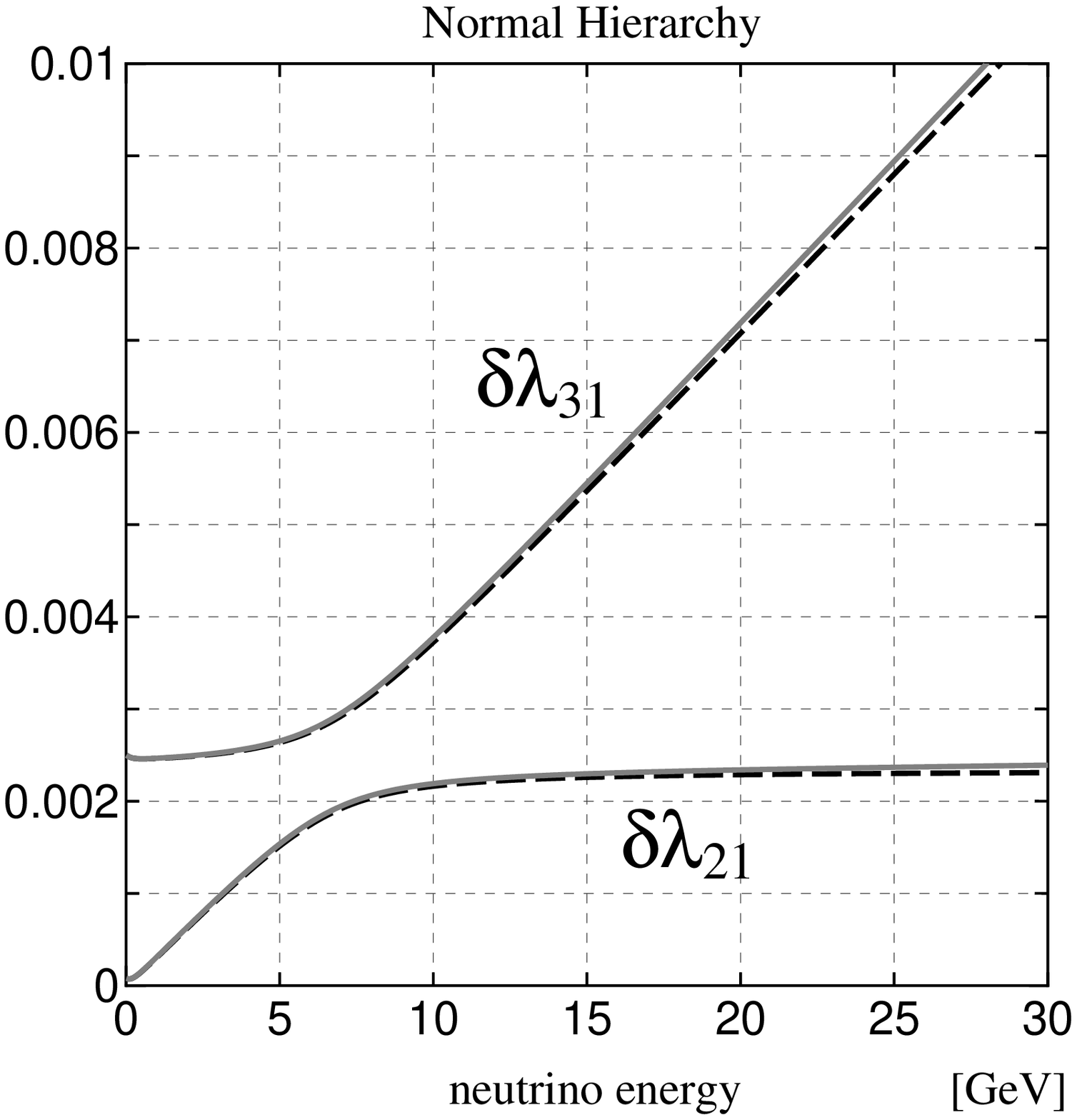}  & 
\includegraphics[scale=0.36]{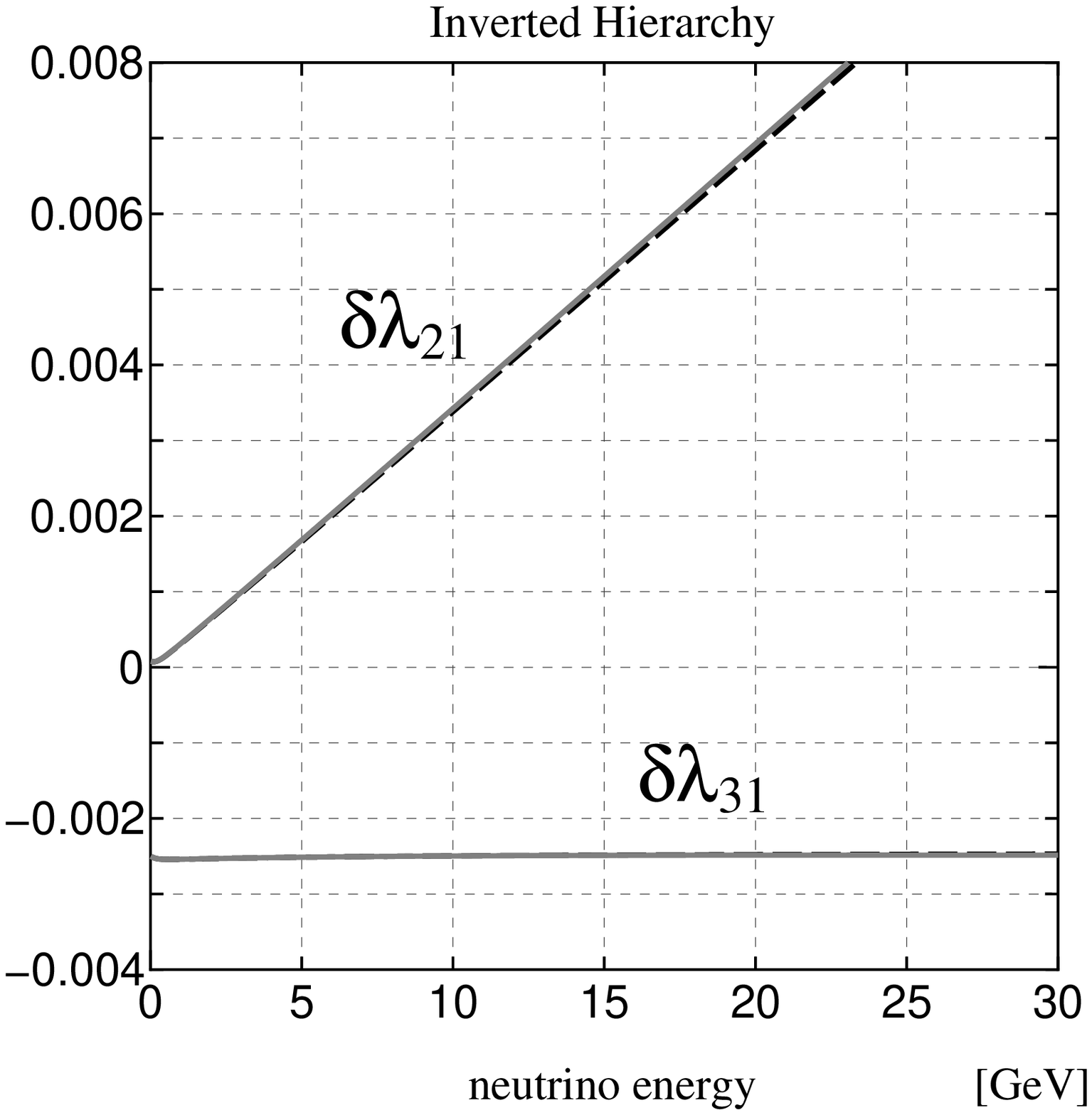}  \\
\includegraphics[scale=0.36]{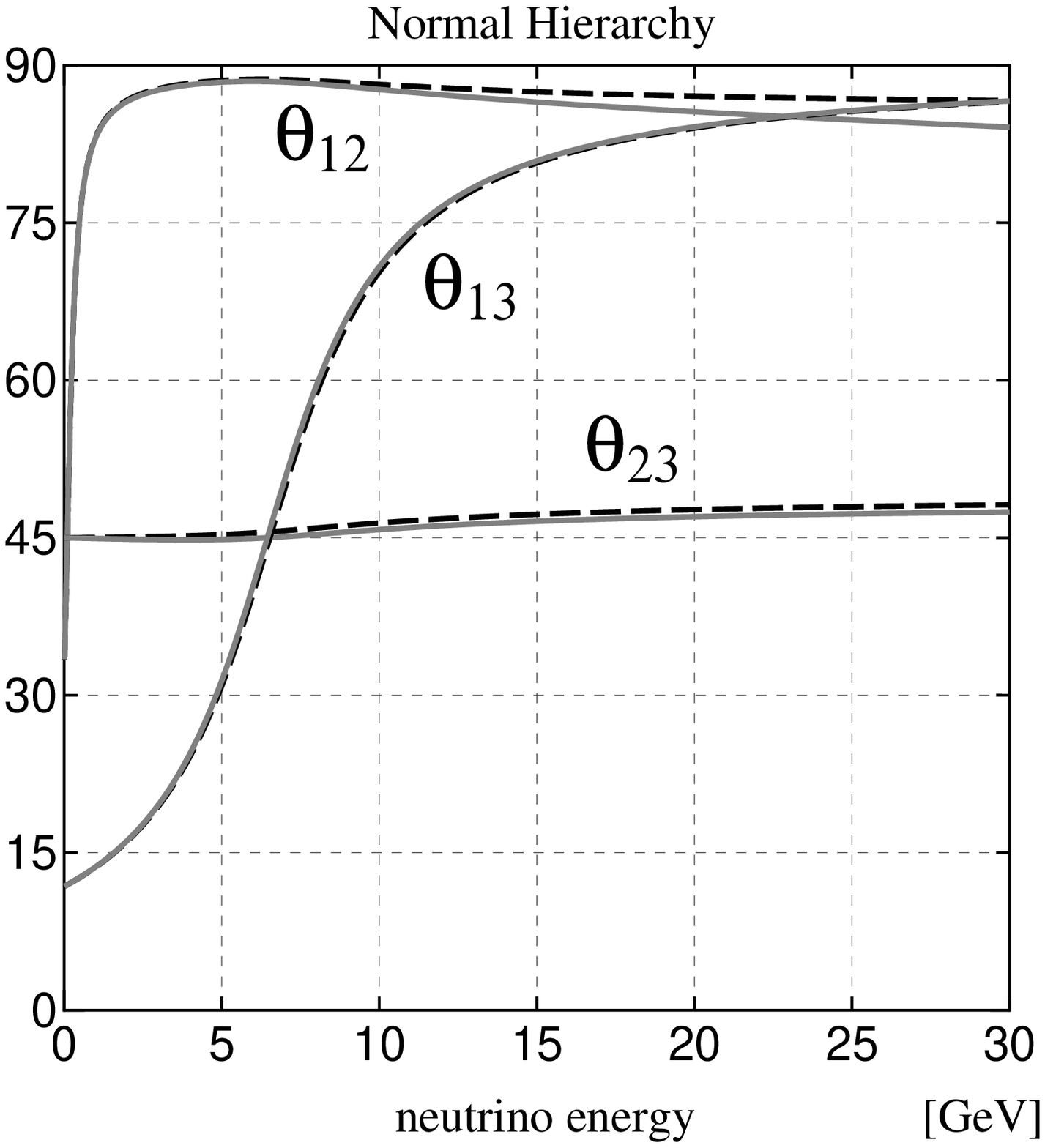} &
\includegraphics[scale=0.36]{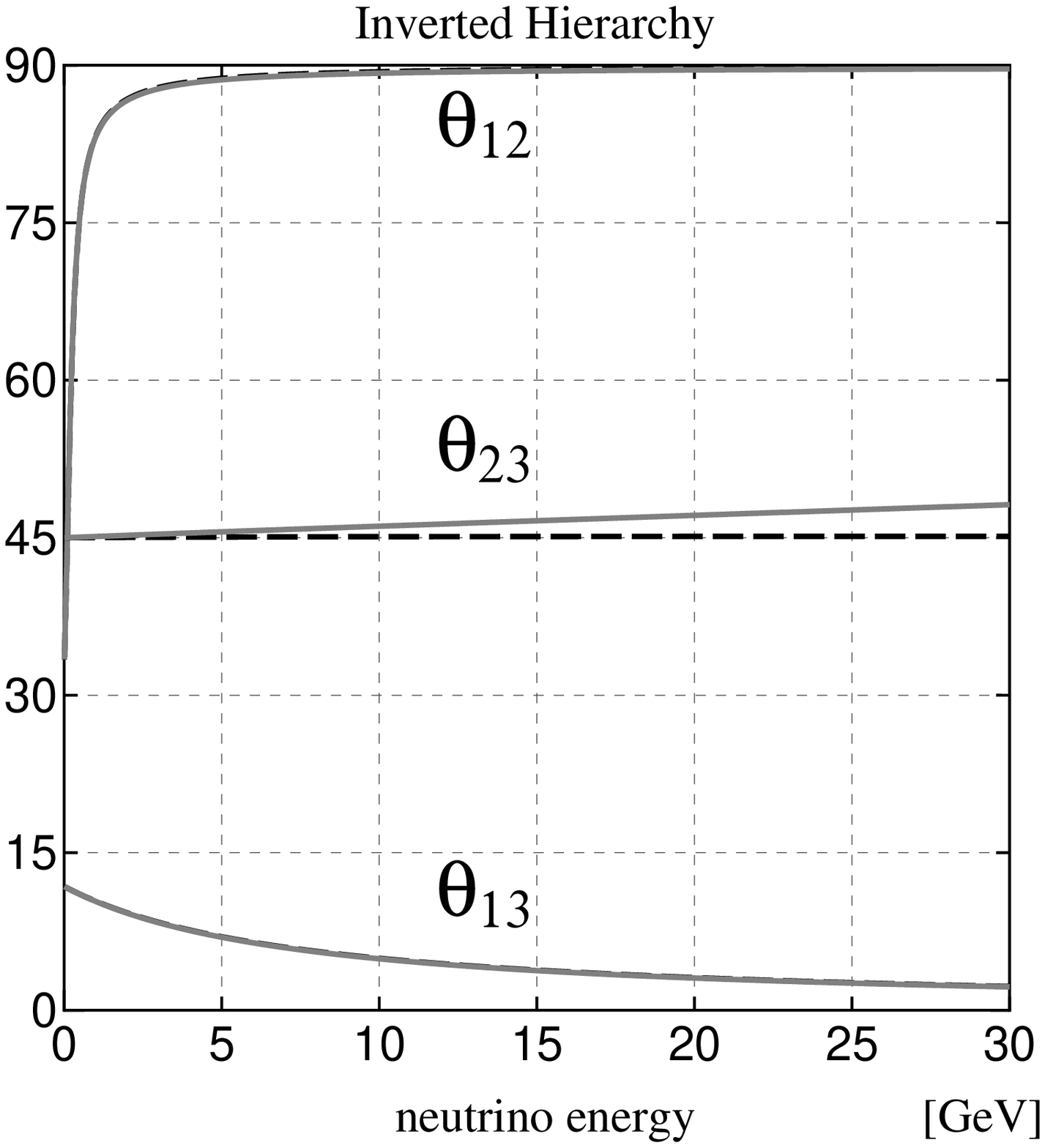} \\
\includegraphics[scale=0.36]{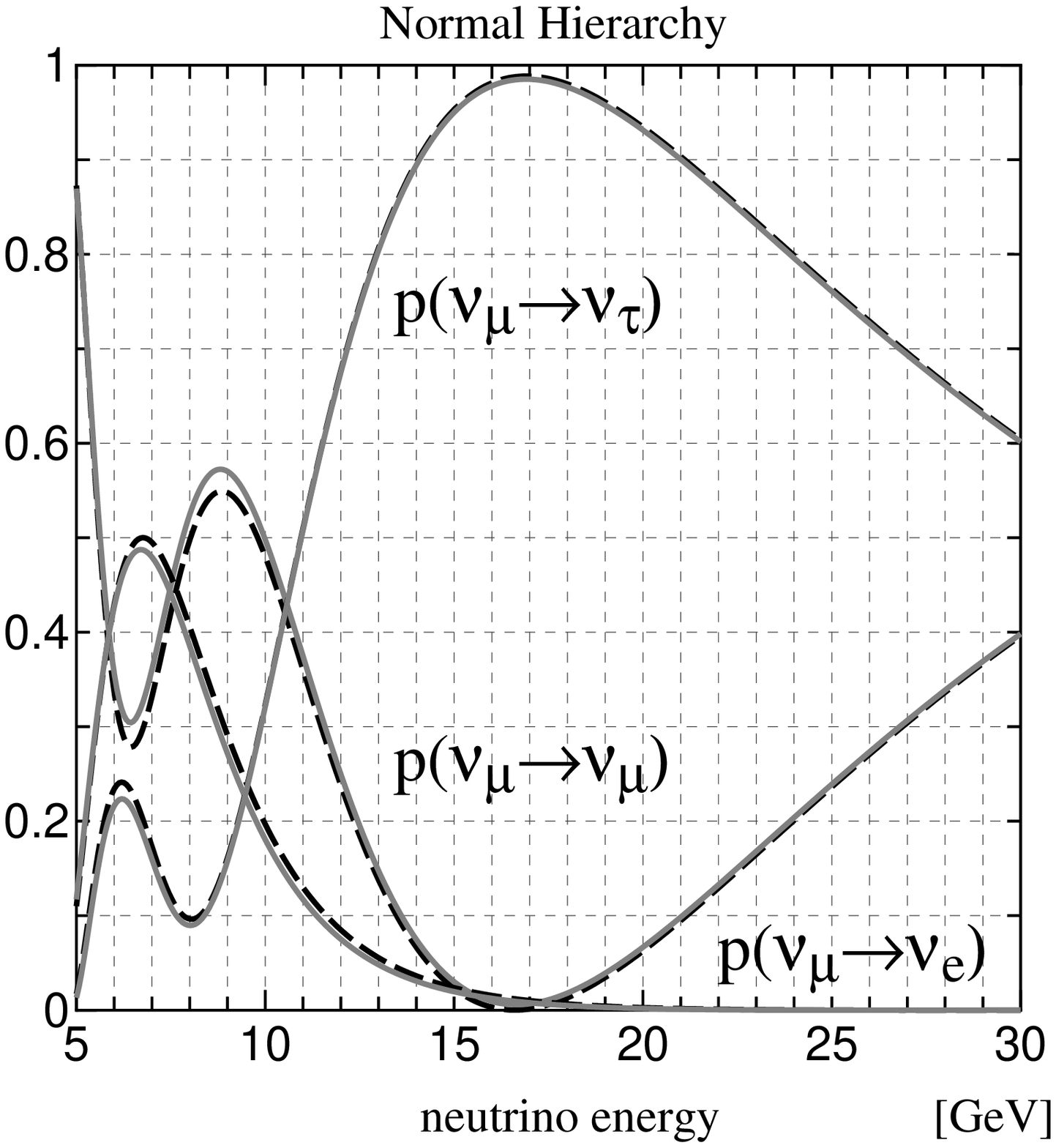}  &
\includegraphics[scale=0.36]{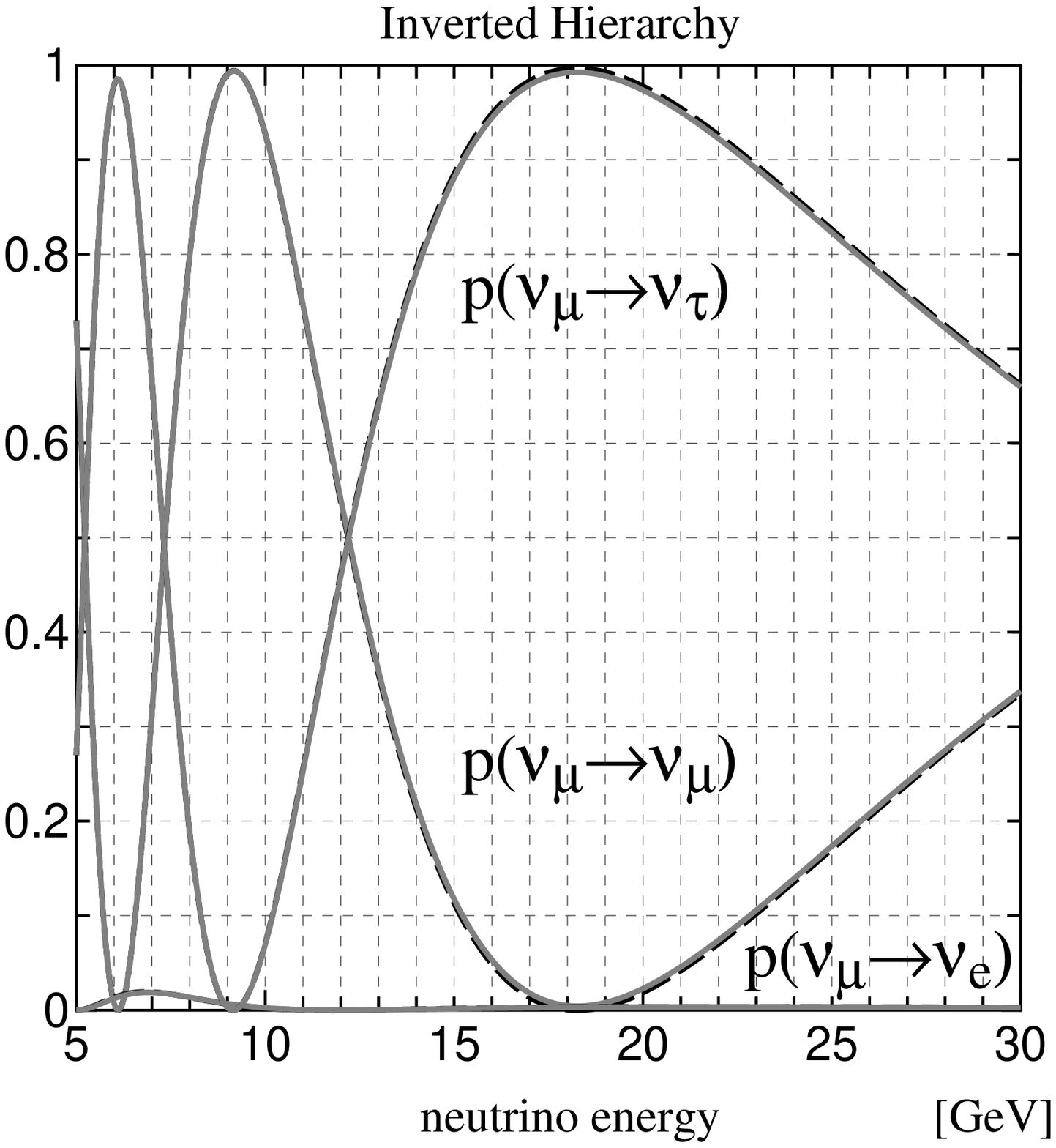} 
\end{array}
\]
\caption{The effective mass-squared-differences, effective mixing angles,
and oscillation probabilities for the case $\sin^2(2\theta_{23})=1$.
The other input parameters are given in Eq.~(\protect{\ref{Inputs}}).
The $\xi=0$ case is plotted with black dashed lines, while the
$\xi=0.025$ case is plotted with gray solid lines.}
\label{Figure1}
\end{center}
\end{figure}
\begin{figure}[p]
\begin{center}
\[
\begin{array}{rr}
\includegraphics[scale=0.36]{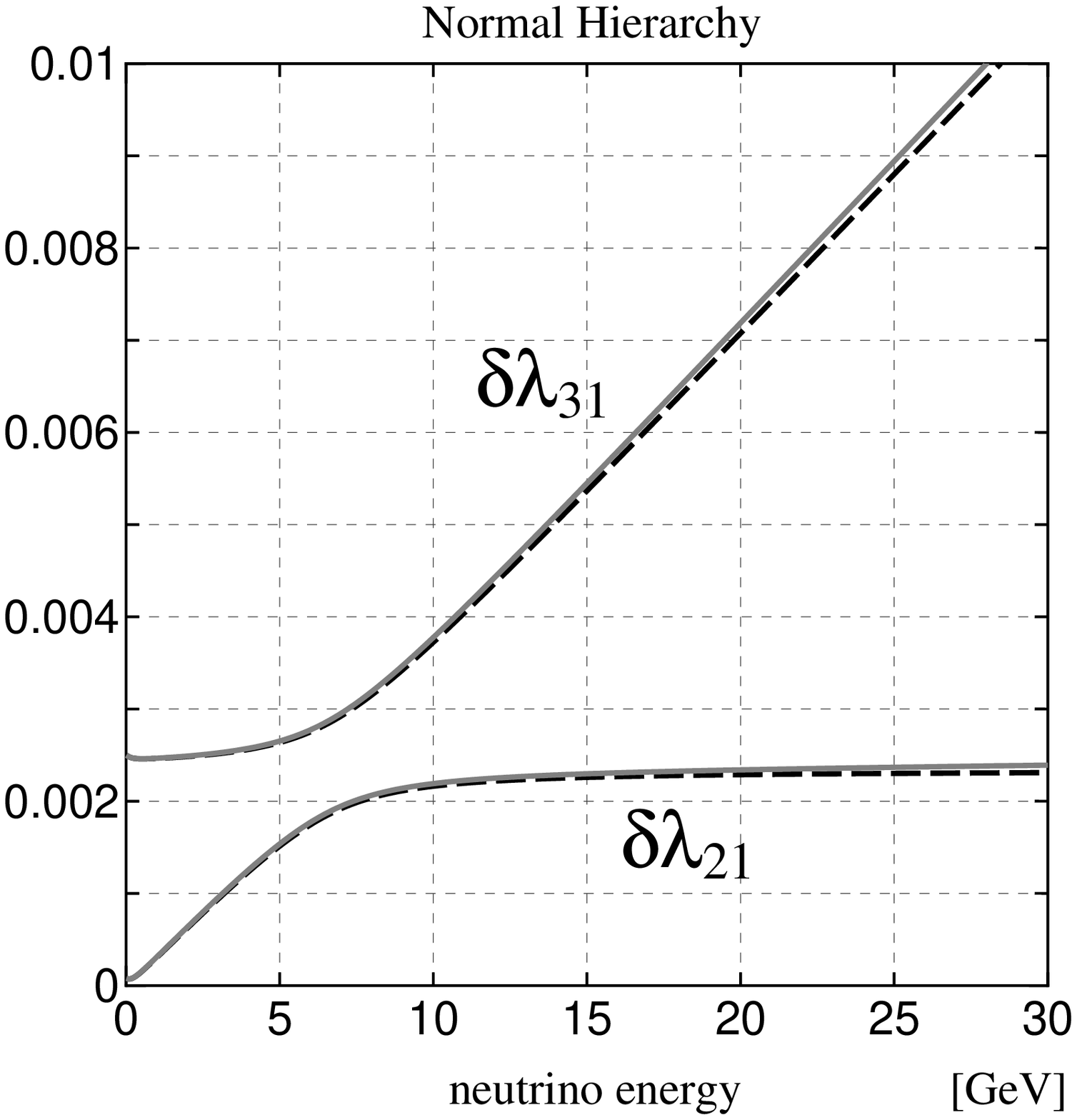}  & 
\includegraphics[scale=0.36]{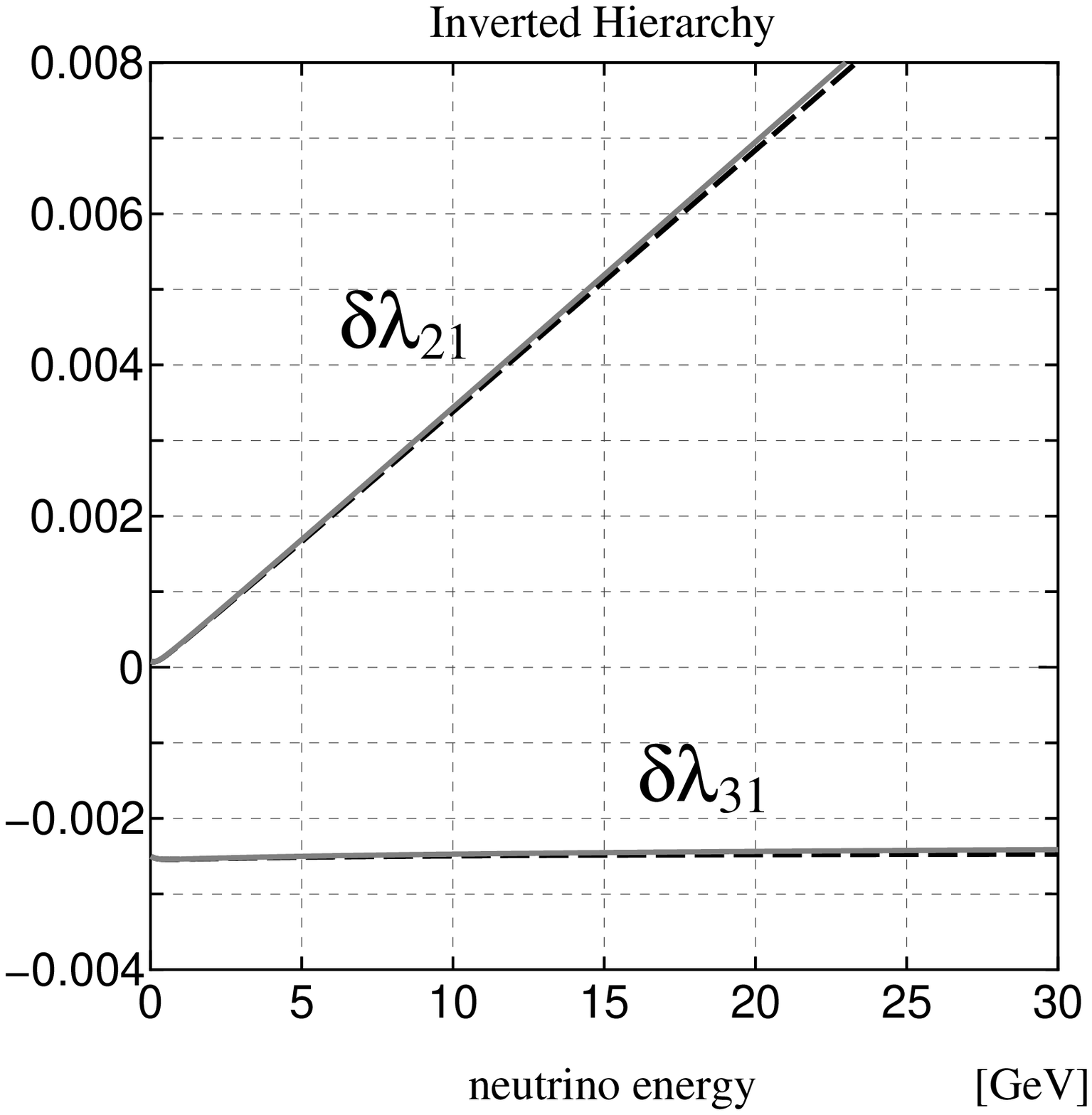}  \\
\includegraphics[scale=0.36]{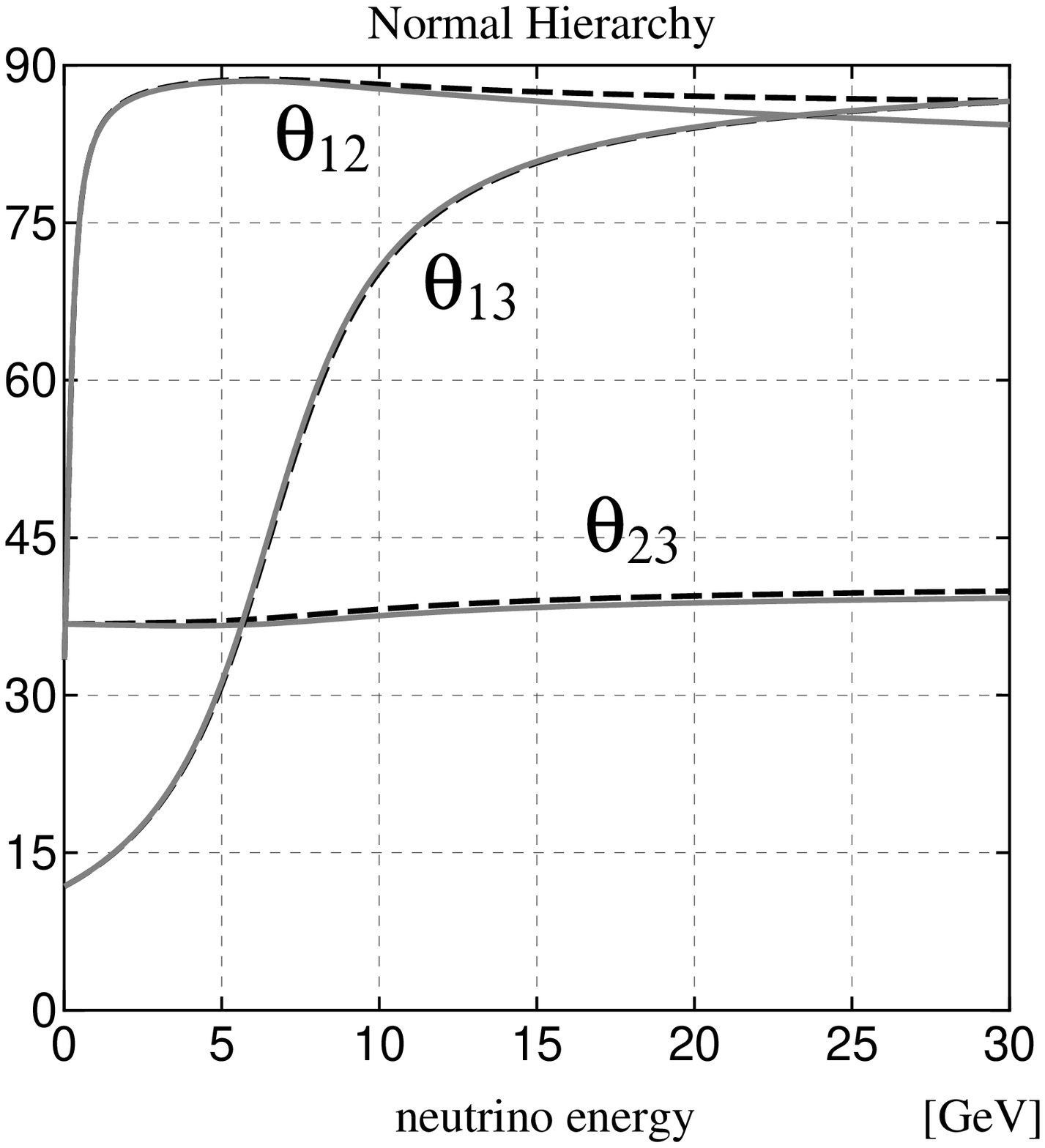} &
\includegraphics[scale=0.36]{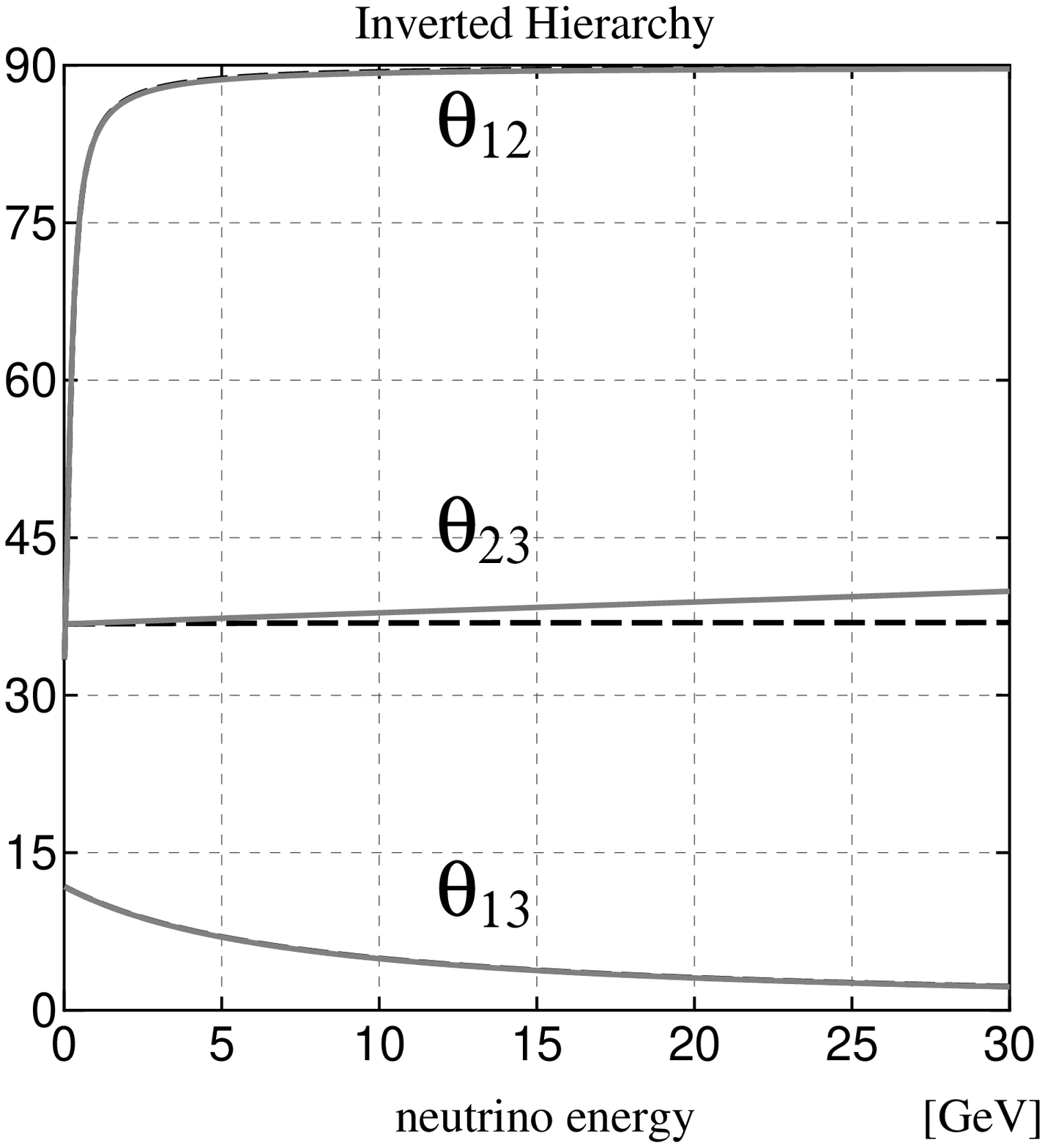} \\
\includegraphics[scale=0.36]{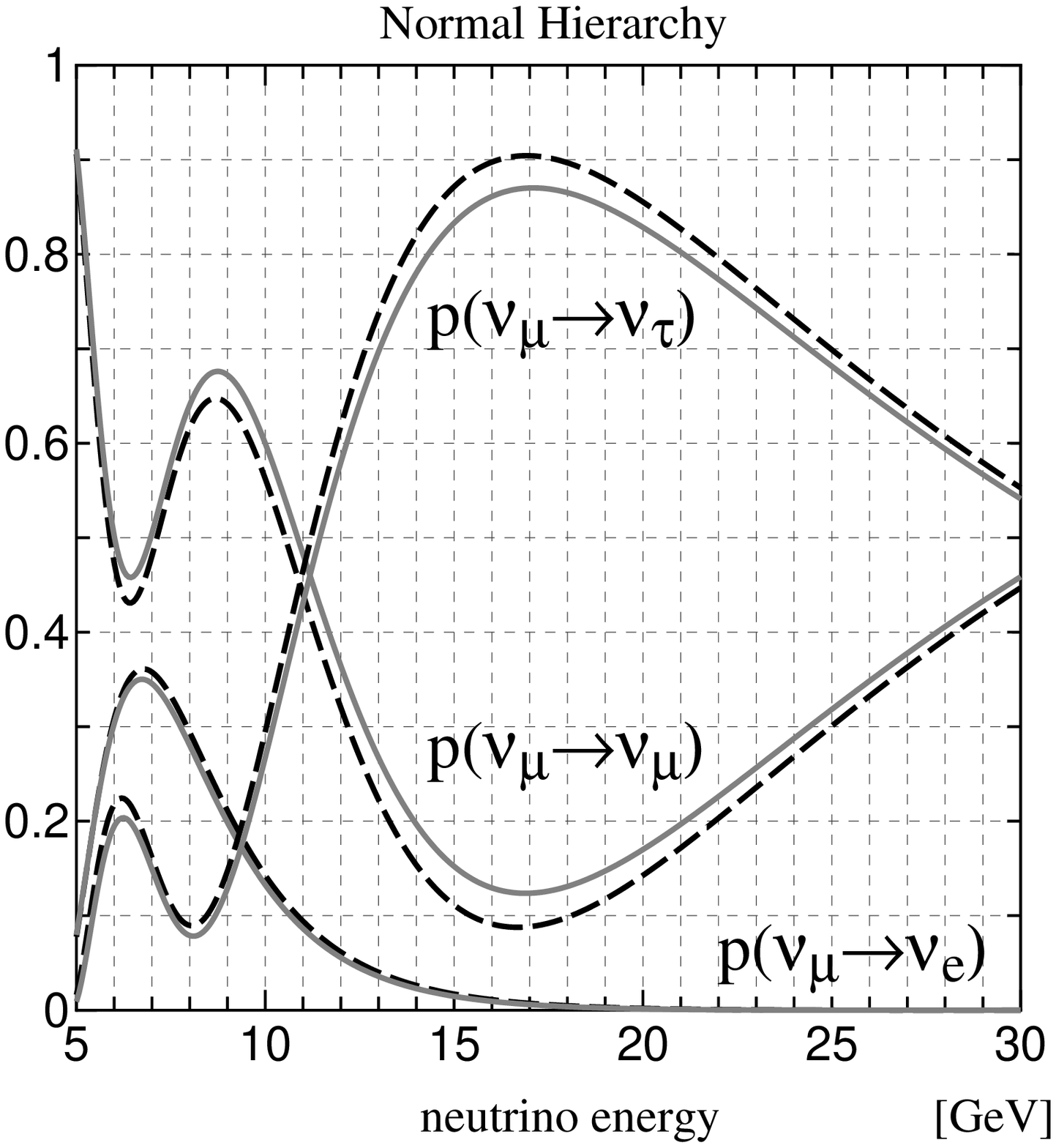}  &
\includegraphics[scale=0.36]{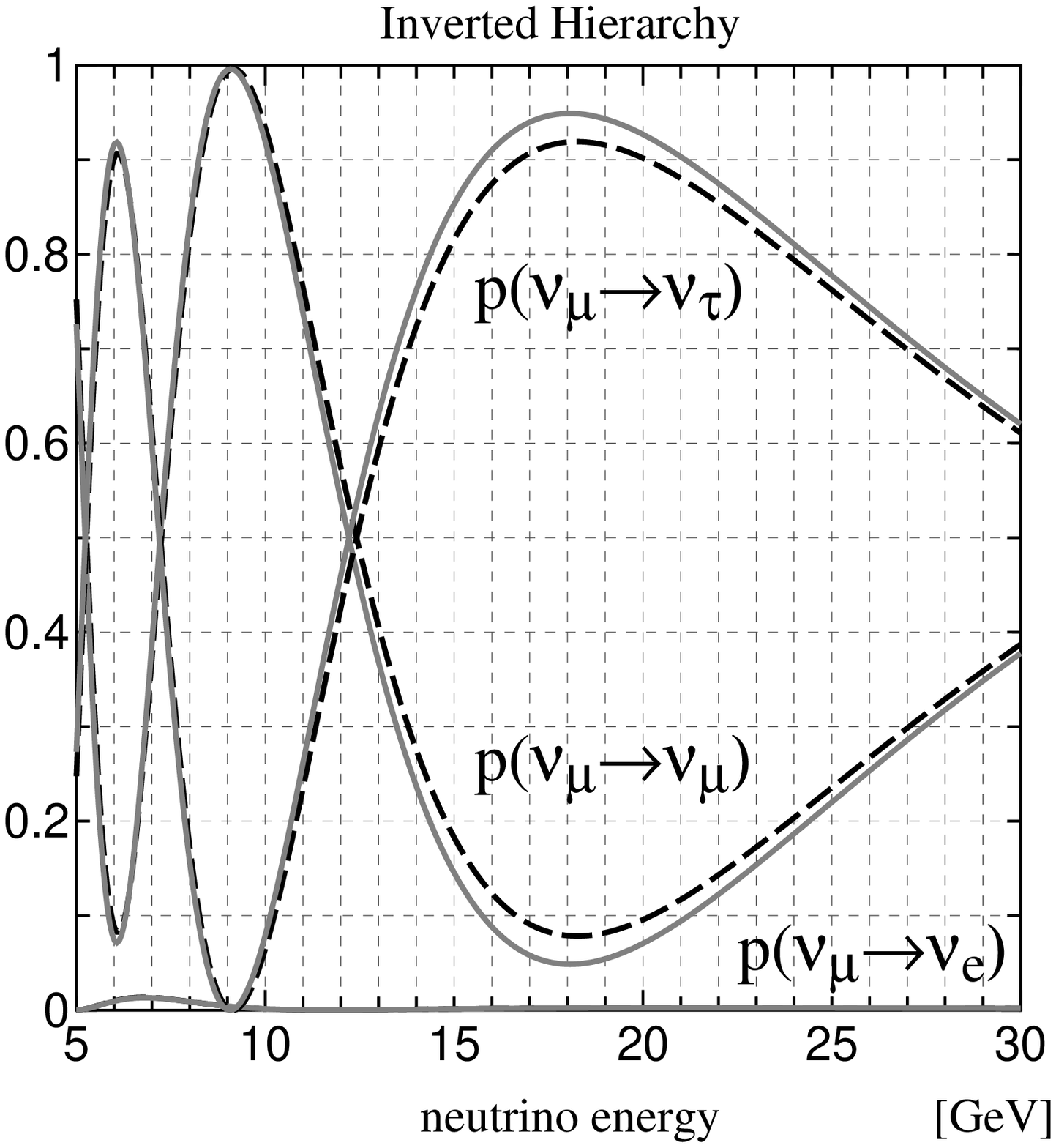} 
\end{array}
\]
\caption{The effective mass-squared-differences, effective mixing angles,
and oscillation probabilities for the case $\sin^2(2\theta_{23})=0.92$
with $\theta_{23}<\pi/4$.
The other input parameters are given in Eq.~(\protect{\ref{Inputs}}).
The $\xi=0$ case is plotted with black dashed lines, while the
$\xi=0.025$ case is plotted with gray solid lines.
}
\label{Figure2}
\end{center}
\end{figure}

As inputs, we use the following:
For $\theta_{23}$, we consider the two cases
\begin{equation}
\sin^2(2\theta_{23})=1\;,\qquad\mbox{and}
\qquad \sin^2(2\theta_{23})=0.92\;\;\mbox{with $\theta_{23}<\dfrac{\pi}{4}$}\;.
\end{equation}
The values of $\theta_{23}$ for these cases are
\begin{equation}
\theta_{23} = \dfrac{\pi}{4}\;,\quad\mbox{and}\quad \theta_{23} = 0.204\,\pi\;.
\end{equation}
For $\xi$, we compare the two cases
\begin{equation}
\xi=0\;,\quad\mbox{and}\quad \xi=0.025\;.
\end{equation}
$\xi=0.025$ corresponds to the central value of the CHARM/CHARM II constraint,
Eq.~(\ref{CHARMconstraint}).
The remaining parameters are fixed to (see Ref.~\cite{HKOT} and references therein) :
\begin{eqnarray}
\delta m^2_{21}      & = & 8.2\times 10^{-5}\,\mathrm{eV}^2\;,\cr
|\delta m^2_{31}|    & = & 2.5\times 10^{-3}\,\mathrm{eV}^2\;,\cr
\tan^2\theta_{12}    & = & 0.4\;,\cr
\sin^2(2\theta_{13}) & = & 0.16\;,\cr
\delta               & = & 0\;,\cr
\rho                 & = & 4.6\,\mathrm{g/cm^3}\;,\cr
L                    & = & 9120\,\mathrm{km}.
\label{Inputs}
\end{eqnarray}
The baseline length of $L=9120\,\mathrm{km}$ is the distance from 
Fermilab to Kamioka, Japan, and the mass density of 
$\rho=4.6\,\mathrm{g/cm^3}$ is the average mass density along this baseline
calculated from the Preliminary Earth Reference Model \cite{PREM}.

Figs.~\ref{Figure1} and \ref{Figure2} show the energy dependence of the effective mass-squared differences, effective mixing angles, and oscillation probabilities of the neutrinos:
Fig.~\ref{Figure1} for the $\sin^2(2\theta_{23})=1$ case, and Fig.~\ref{Figure2} for the
$\sin^2(2\theta_{23})=0.92\;(\theta_{23}<\pi/4)$ case. 
In the figures, the $\xi=0$ case is plotted in broken black lines
while the $\xi=0.025$ case is plotted in solid gray lines.
As can be clearly seen from the graphs in the top rows of both figures, 
the effective mass-squared differences are little affected by $\xi$ as expected.
On the other hand, the graphs in the middle rows show
that of the effective mixing angles, $\tilde{\theta}_{12}$ is shifted in the negative direction 
when $\delta m^2_{31}>0$ (normal hierarchy), 
while $\tilde{\theta}_{23}$ is shifted in the positive direction
when $\delta m^2_{31}<0$ (inverted hierarchy), as tabulated in Table.~\ref{EffectOfNCUV}.
However, the graphs on the bottom rows show that these shifts in the mixing angles
are virtually invisible in the oscillation probabilities when $\sin^2(2\theta_{23})=1$,
but quite visible when $\sin^2(2\theta_{23})=0.92$, again as expected.

Numerical calculations for the anti-neutrino case also confirm the accuracy of our
analytical results, though we will not present them here.

\section{Fermilab $\rightarrow$ Hyper-Kamiokande}

If the value of $\sin^2(2\theta_{23})$ is not too close to one, 
then matter effects due to neutral current universality violation
will lead to shifts in the oscillation probabilities,
as shown in the bottom row of Fig.~\ref{Figure2}.
Let us now ask whether such shifts are observable in 
long baseline neutrino oscillation experiments.
In the following, we will assume that $\sin^2(2\theta_{23})=0.92\;(\theta_{23}<\pi/4)$, 
which was the value used in Fig.~\ref{Figure2}, and that it is accurately known.

The effect we would like to see only appears at mass densities and energies at
which
\begin{equation}
|\delta m^2_{31}| < a 
= \left( 7.6324\times 10^{-5}\mathrm{eV}^2 \right)
\times\left(\dfrac{\rho}{\mathrm{g/cm^3}}\right)
\times\left(\dfrac{E}{\mathrm{GeV}}\right) \;,
\end{equation}
or
\begin{equation}
(26\sim 39) <
\left(\dfrac{\rho}{\mathrm{g/cm^3}}\right)
\times\left(\dfrac{E}{\mathrm{GeV}}\right) \;,
\end{equation}
for $|\delta m^2_{31}|=(2\sim 3)\times 10^{-3}\mathrm{eV}^2$.
Since the mass density of the Earth's crust and mantle are
$3\sim 5\,\mathrm{g/cm^3}$ \cite{PREM}, this requires the beam energy to be larger than
$\sim 10\,\mathrm{GeV}$.   
At these energies, the position of the first oscillation peak (dip) is
determined by the condition
\begin{equation}
\dfrac{a}{2E}L \sim \pi \;,
\end{equation}
which translates to
\begin{equation}
\left(\frac{\rho}{\mathrm{g/cm^3}}\right)\times
\left(\frac{L}{\mathrm{km}}\right) \sim \;3\times 10^4\;,
\end{equation}
or
\begin{equation}
L \sim 10^4\,\mathrm{km}\;.
\end{equation}
Therefore, the experiment we need to consider is such that a neutrino
beam of energy in excess of 10~GeV is aimed at a detector about
10,000~km away.

\begin{figure}[t]
\includegraphics[scale=0.7]{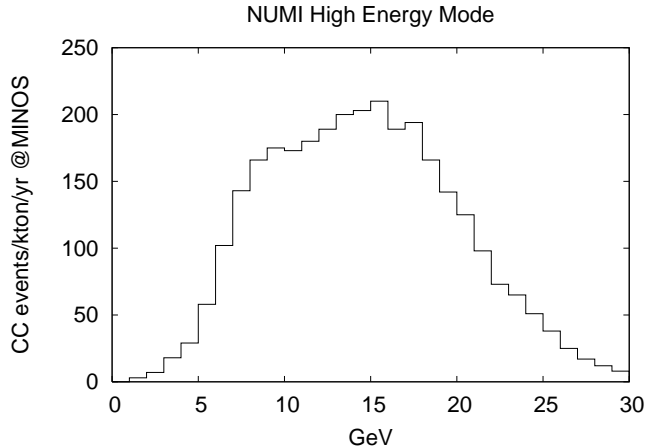}
\caption{The NUMI beam in its high energy mode. 
The vertical axis gives the number of expected charged current $\nu_\mu$ events
at the MINOS site (732~km away from Fermilab) per kiloton of detector material, per year, per GeV bin
without any oscillation.}
\label{NUMIprofile}
\end{figure}

At this point, we note that a $\nu_\mu$ beam with the required energies
is already available at Fermilab.
Fig.~\ref{NUMIprofile} is reproduced from the 
NUMI Technical Design Handbook \cite{NUMI} and shows 
the energy profile of the NUMI beam in its high energy mode.
As we can see, the beam has considerable support in the $5\sim 25$~GeV range.
The vertical axis is the expected number of charged current $\nu_\mu$ events at MINOS 
per kiloton of detector material, per year, per GeV bin without any oscillation.
If a similar beam were aimed at a detector $\sim 10^4\,\mathrm{km}$ away, which is more
than 10 times the distance from Fermilab to MINOS, the $\nu_\mu$ flux will be
attenuated by at least 2 orders of magnitude from what is available at MINOS.
Therefore, a megaton class detector would be required if the number of observed
events is to be statistically significant.

\begin{figure}[p]
\includegraphics[scale=0.8]{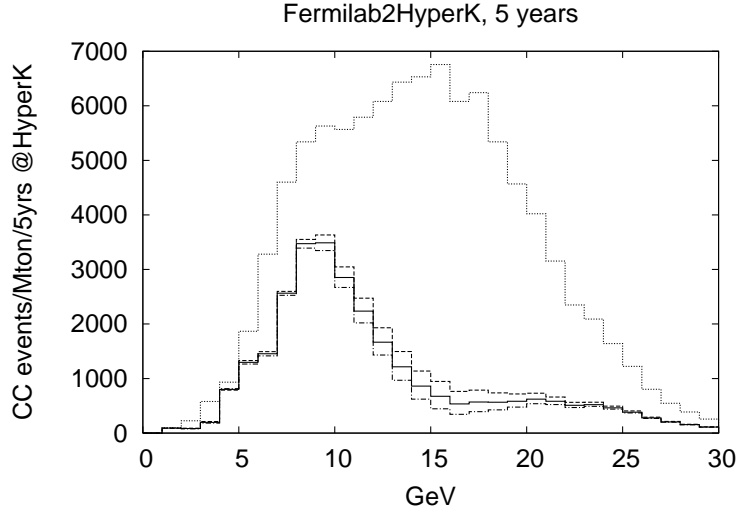}
\caption{The expected number of $\nu_\mu$ events at a Fermilab$\rightarrow$HyperK experiment with
 5 years of data taking.
The dotted line indicates the number of events without any oscillation.
The solid, dashed, and dot-dashed lines indicate the number of expected events with oscillation taken into account for the $\xi=0$, $\xi=+0.025$, and $\xi=-0.025$ cases, respectively. The mass hierarchy assumed was the normal hierarchy, $\sin^2(2\theta_{23})=0.92\;(\theta_{23}<\pi/4)$, and the other input
parameters were those listed in Eq.~(\protect{\ref{Inputs}}).}
\label{Fermilab2HyperK1}
\end{figure}
\begin{figure}[p]
\includegraphics[scale=0.8]{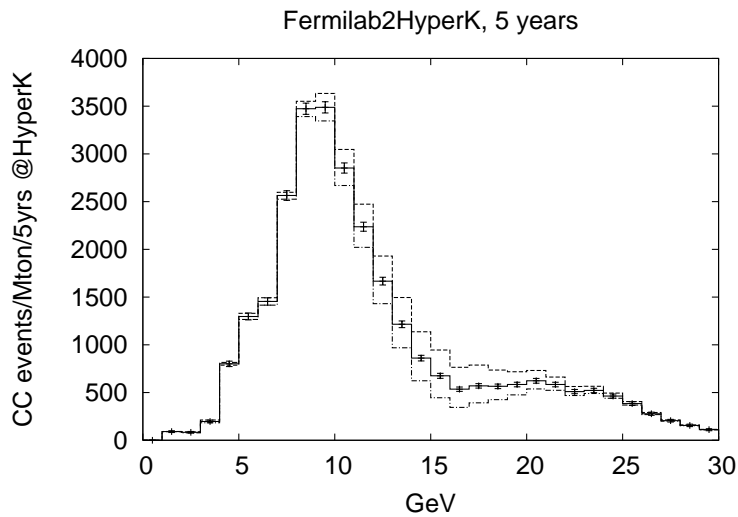}
\caption{Blowup of Fig.~\protect{\ref{Fermilab2HyperK1}}.
Error bars have been added for the $\xi=0$ case.}
\label{Fermilab2HyperK2}
\end{figure}

The planned Hyper-Kamiokande (HyperK) \cite{HyperK} is a megaton 
water-Chrenkov detector which would be at a distance of $L=9120\,\mathrm{km}$ from Fermilab.
Aiming a NUMI-like beam from Fermilab at HyperK (the declination angle is 46 degrees)
would provide the necessary energy, detector mass, and baseline length.
So this is the setup we will consider.
The average matter density along the baseline would be $4.6\,\mathrm{g/cm^3}$,
and the oscillation probability to be measured will be that shown
in Fig.~\ref{Figure2}.

In Fig.~\ref{Fermilab2HyperK1}, we show the expected number of $\nu_\mu$ events
at HyperK for 5 years of data taking.  The dotted line indicates the expected
numbers without any oscillation, and was obtained by rescaling the numbers from
Fig.~\ref{NUMIprofile} to take into account the difference in baseline length,
detector mass, and the number of years of data taking.
The solid line indicates the expected number of events with oscillation
taken into account for the normal hierarchy case with $\xi=0$.
The input parameters are those listed in Eq.~(\ref{Inputs}) with 
$\sin^2(2\theta_{23})=0.92\;(\theta_{23}<\pi/4)$.
The $\xi=+0.025$ and $\xi=-0.025$ cases are shown with dashed, and dot-dashed lines,
respectively.
As one can see from the figure, the expected
number if events is fairly large even at this distance, and even with oscillation.

Fig.~\ref{Fermilab2HyperK2} shows a blowup of Fig.~\ref{Fermilab2HyperK1}.
Even at the oscillation dip at 17~GeV, the expected number of events is in the hundreds.
Due to this significant statistics,
the $\xi=0$ and $\xi=\pm 0.025$ cases are clearly distinguishable as indicated by
the error bars on the $\xi=0$ case.  Therefore, this experiment can easily
detect a violation in neutral current universality if it is as large as the
CHARM/CHARM II central value.

\begin{figure}[t]
\includegraphics[scale=0.8]{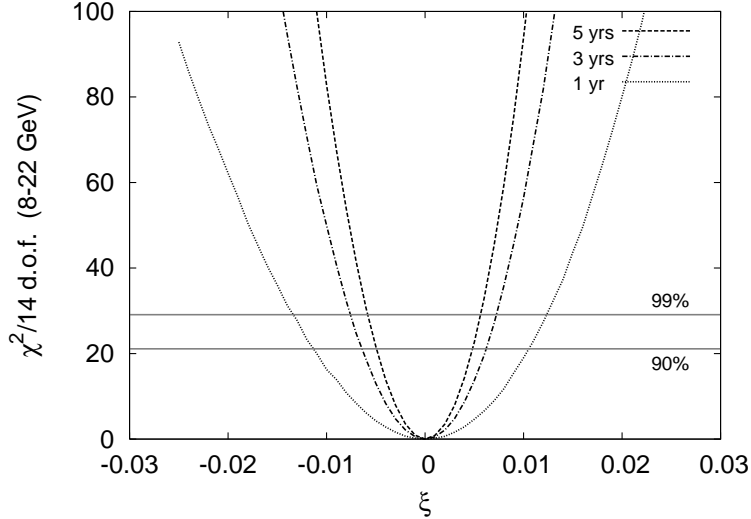}
\caption{The $\chi^2$ for the 14 bins from 8 to 22 GeV between the $\xi=0$ and $\xi\neq 0$ cases.
}
\label{chi2-14dof}
\end{figure}

To see what kind of constraint this experiment could place on $\xi$, we calculate
the $\chi^2$ between the $\xi=0$ and the $\xi\neq 0$ cases, \textit{i.e.}
\begin{equation}
\chi^2(\xi) \equiv \sum_{8\,\mathrm{GeV}<i<22\,\mathrm{GeV}}
\dfrac{ [\,N_i(\xi)-N_i(0)\,]^2 }{ N_i(0) }\;,
\end{equation}
where $N_i(\xi)$ is the expected number of events in the $i$-th GeV-wide bin,
and plot the $\xi$-dependence of the $\chi^2(\xi)$ in Fig.~\ref{chi2-14dof}.
We have restricted the bins that enter into $\chi^2(\xi)$ to the 8 to 22 GeV range 
(14 bins), since
that is the range in which the expected number of events fluctuates significantly with $\xi$.
With 5 years of data taking, we can read off from the graph that the $\xi=0$ and $\xi=\pm 0.005$
cases are distinguishable at the 99\% confidence level.
This corresponds to a limit on universality violation at the $1\%$ level, which 
will be comparable to
the constraints from LEP/SLD \cite{Lebedev:1999vc,Loinaz:2004qc,Chang:2000xy}
but completely model independent.
For fewer years of data taking, the limits will be correspondingly weaker.

We emphasize that the conclusions in this section are valid only for the 
$\sin^2(2\theta_{23})=0.92 \;(\theta_{23}<\pi/4)$ case.
The closer $\sin^2(2\theta_{23})$ is to one, the more difficult it will be to detect the
presence or absence of $\xi$.

\section{Summary and Conclusions}

In this paper, we have considered the matter effect on neutrino oscillations due to
neutral current universality violation.   It was shown that the effect of the violation
appears dominantly as a shift in the effective value of $\theta_{23}$ at high energies,
while the other effective mixing angles and effective mass-squared-differences are virtually unaffected.
As a result, the effect will manifest itself as changes in the amplitudes of the oscillation probabilities, 
while the locations of the oscillation peaks and dips in distance/energy remain the same.
However, since the amplitudes of the $\nu_\mu$ and $\bar{\nu}_\mu$ survival probabilities are
proportional to $\sin^2(2\theta_{23})$, the shift in $\theta_{23}$ would be difficult to
detect if $\sin^2(2\theta_{23})$ is too close to one.

If the value of $\sin^2(2\theta_{23})$ is as small as 0.92, the current 90\% lower bound, 
then a 5-year measurement of the $\nu_\mu$ survival spectrum 
by a Fermilab$\rightarrow$HyperK experiment could place a model-independent 
constraint on neutral current universality violation at the 1\% level.
This would be competitive with the model-dependent constraints extracted from LEP/SLD data
\cite{Lebedev:1999vc,Loinaz:2004qc,Chang:2000xy}.

The analysis in this paper was restricted to the $\delta =0$ case, in which the
effective $\theta_{23}$ was unaffected by charged-current interactions.  
For the $\delta\neq 0$ cases, one needs to
account for the charged-current shift discussed in Ref.~\cite{HKOT}, 
in addition to the neutral-current shift discussed in this paper, 
making the analysis somewhat more complicated.
However, for the neutrino case with inverted hierarchy ($\delta m^2_{31}<0$),
and the anti-neutrino case with normal hierarchy ($\delta m^2_{31}>0$),
charged-current effects are always absent from $\theta_{23}$, regardless of the value of $\delta$.
Therefore, using neutrinos if the hierarchy is inverted, and anti-neutrinos if the
hierarchy is normal, can potentially provide a clean signal.

\section*{Acknowledgments}

We would like to thank Hiroaki Sugiyama for helpful discussions.
Takeuchi would like to thank the hospitality of the particle theory group at
Ochanomizu Women's University,
where a major portion of this work was carried out during the summer of 2005.
Numerical calculations were performed on Altix3700 BX2 at the
Yukawa Institute of Theoretical Physics at Kyoto University.
This research was supported in part by the U.S. Department of Energy, 
grant DE--FG05--92ER40709, Task A (T.T.).


\end{document}